%% file: ANA-TOPQ-2020-10-PAPER.tex
\pdfoutput=1
\documentclass[cernpreprint, atlasdraft=false, texlive=2016, UKenglish, texmf]{atlasdoc}
 
\usepackage{atlaspackage}
\usepackage[backref=false]{atlasbiblatex}
 
\usepackage[hepparticle]{atlasphysics}
 
\usepackage[output=true]{atlastodo}
\ATLnote{SB}{Simon}{orange!20}{red!50!black!60}
\ATLnote{Q}{Quake}{green!20}{red!50!black!60}
\ATLnote{LS}{Lisa}{yellow!20}{red!50!black!60}
\ATLnote{CN}{Clara}{blue!20}{red!50!black!60}
 
\graphicspath{{logos/}{figures/}}
 
\usepackage{ANA-TOPQ-2020-10-PAPER-defs}

\AtlasTitle{Measurement of the \tttt{} production cross section in $pp$ collisions at $\sqrt{s}=13$~\tev\ with the ATLAS detector}

\AtlasAbstract{
A measurement of four-top-quark production using proton--proton collision data
at a centre-of-mass energy of $13~\tev$ collected by the ATLAS detector
at the Large Hadron Collider corresponding to an integrated luminosity of 139~fb\(^{-1}\) is presented.
Events are selected if they contain a single lepton (electron or muon) or an opposite-sign lepton pair,
in association with multiple jets.
The events are categorised according to the number of jets and how likely these
are to contain $b$-hadrons. A multivariate technique is then used to discriminate
between signal and background events.
The measured four-top-quark production cross section
is found to be $26^{+17}_{-15}$~fb,
with a corresponding observed (expected) significance of 1.9 (1.0) standard deviations
over the background-only hypothesis.
The result is combined with the previous measurement performed by the ATLAS Collaboration
in the multilepton final state.
The combined four-top-quark production cross section is measured to be $24^{+7}_{-6}$~fb,
with a corresponding observed (expected) signal significance of 4.7 (2.6) standard deviations
over the background-only predictions.
It is consistent within 2.0 standard deviations with the Standard Model
expectation of $12.0\pm2.4$~fb.
}
 
\author{The ATLAS Collaboration}
 
\AtlasRefCode{TOPQ-2020-10}
 
\PreprintIdNumber{CERN-EP-2021-075}

\AtlasJournalRef{JHEP 11 (2021) 118}
\AtlasDOI{10.1007/JHEP11(2021)118}

\hypersetup{pdftitle={ATLAS document},pdfauthor={The ATLAS Collaboration}}
 
\begin{document}
 
\maketitle
 
\tableofcontents
 
% The next lines are included from the .//intro/intro.tex input file
\section{Introduction}
\label{sec:introduction}
 
\begin{figure}[!htb]
\centering
\includegraphics[width=0.32\textwidth]{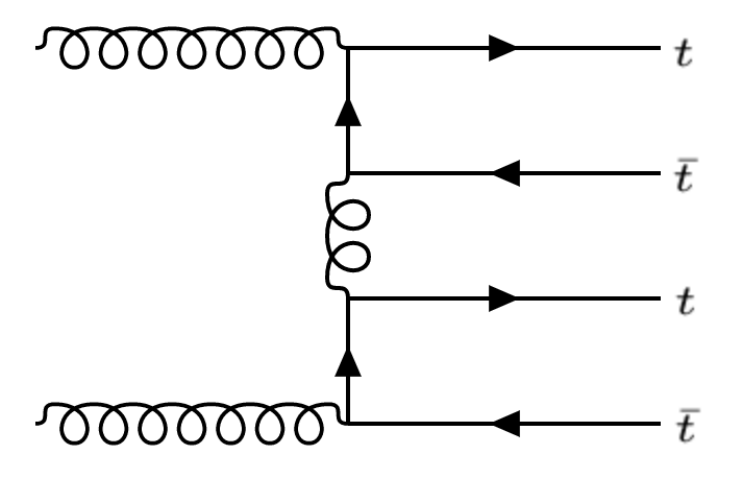}
\includegraphics[width=0.32\textwidth]{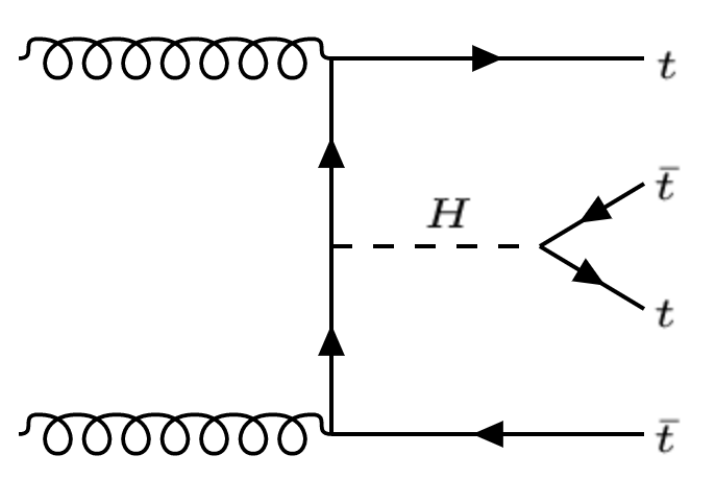}
\includegraphics[width=0.32\textwidth]{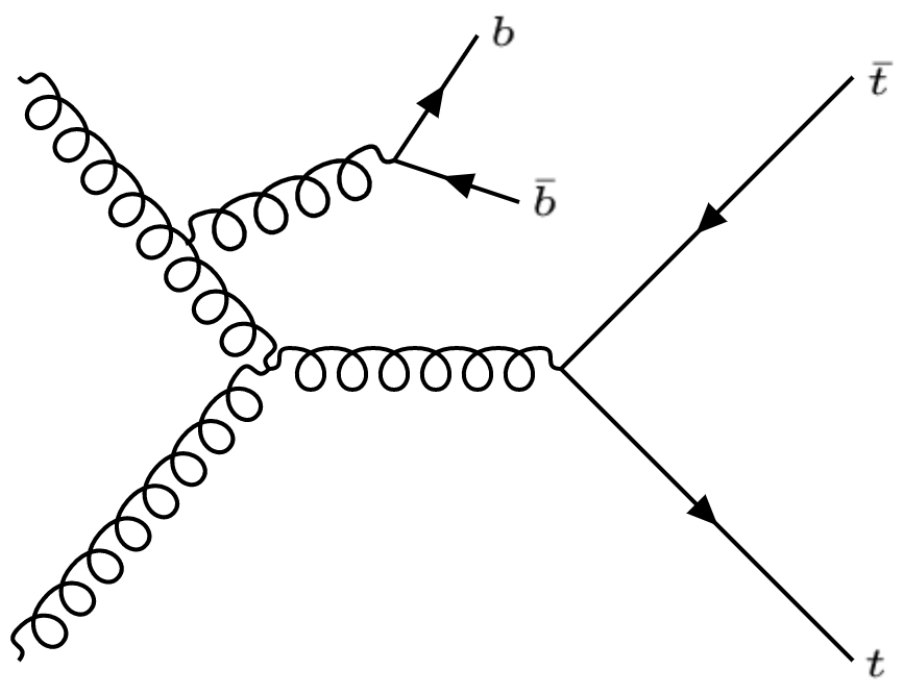}
\caption{Examples of tree-level Feynman diagrams for SM \tttt signal (left and middle) and one
of the main backgrounds, $\ttbar$ production in association with $b$-jets (right).}
\label{fig:feynman}
\end{figure}

Being the heaviest known elementary particle of the Standard Model (SM),
the top quark has a large coupling to the SM Higgs boson
and is predicted to have large couplings to hypothetical new particles in many models
of physics beyond the Standard Model.
For this reason, it is particularly important to study rare processes involving the top quark.
The production of four top quarks, \tttt, is one of these processes which has not yet been observed.
The \tttt\ cross section could be enhanced, for instance, by
gluino pair production as in supersymmetric theories~\cite{Nilles:1983ge,Farrar:1978xj},
by pair production of scalar gluons~\cite{Plehn:2008ae,Calvet:2012rk},
or by the production of a heavy scalar or pseudoscalar boson in association with
a top-quark pair (\ttbar) in type-II two-Higgs-doublet models~\cite{Dicus:1994bm,Craig:2015jba,Craig:2016ygr}.
The \tttt cross section is also sensitive to both the magnitude and the charge conjugation and parity properties of the Yukawa
coupling of the top quark to the Higgs boson~\cite{Cao:2017,Cao:2019ygh},
as well as to various four-fermion couplings in the context of the effective field theory
framework~\cite{Degrande:2010kt,Zhang:2017mls}.
Within the SM, the \tttt cross section in proton--proton (\pp) collisions at a
centre-of-mass energy
of $\sqrt{s}=13$~\tev\ is predicted to be $\stttt = 12.0~\text{fb}$
at next-to-leading order (NLO) in QCD including NLO electroweak
corrections~\cite{Frederix:2017wme}, with a relative combined uncertainty of 20\%
dominated by the missing higher order QCD correction
evaluated by varying the renormalisation and factorisation scale choices.
The uncertainty from the PDF and \alphas choice in the \tttt cross section
was estimated to be about 6.3\% using the PDF4LHC prescription.
An example of a Feynman diagram for SM \tttt\ QCD production is shown in Figure~\ref{fig:feynman} (left).
The electroweak \tttt contribution is illustrated in Figure~\ref{fig:feynman} (middle) with an example of
a Feynman diagram where a Higgs boson acts as an off-shell mediator.
 
The \tttt events can give rise to several different final states
depending on the hadronic or semileptonic decay mode of each of the top quarks.
The final states can be grouped according to the number of electrons or muons from the semileptonic top-quark decays,
including those from subsequent leptonic $\tau$ decays.
The final state with two leptons\footnote{In the rest of this article, `lepton' refers exclusively to an electron or muon.}
of the same electric charge or with more than two leptons is referred to as the \SSML final state.
This final state contains 13\% of all produced \tttt events and features a low level of background contamination.
The final state with one lepton or two oppositely charged leptons (referred to as the \SLOS final state)
accounts for a larger fraction, capturing 57\% of all produced \tttt events.
However, this final state suffers from a large irreducible background that is mostly composed of \ttbar production in association with additional jets (\ttjets).
It is complementary to the \SSML final state and poses different challenges.
The main challenge lies in the proper
evaluation of the dominant background from $t\bar{t}b\bar{b}$ events with
additional jets, which has significant theoretical uncertainty.
An example Feynman diagram for this background is shown in Figure~\ref{fig:feynman} (right).
 
ATLAS and CMS have already searched for \tttt production in 13~\tev\ \pp collisions
at the Large Hadron Collider (LHC).
The most recent ATLAS result focused on the \SSML\ final state using 139~\ifb\ of data and led to the
first evidence for this process with
an observed (expected) significance of 4.3 (2.4) standard deviations
and a measured cross section of $\stttt=24^{+7}_{-5}$~fb~\cite{Aad:2020klt}.
The previous ATLAS search in the \SLOS\ final state using 36.1~\ifb\
set an observed (expected)
95\% confidence level (CL) upper limit on \stttt of
47~fb (33~fb)~\cite{EXOT-2017-11}.
CMS also set a 95\% CL upper limit on the \tttt production cross section in this final state
of 48~fb using a 35.8~\ifb\ data set~\cite{Sirunyan:2019nxl}.
The latest CMS search with 137~\ifb\ of data in the \SSML\ final state lead to an observed (expected)
significance for a \tttt\ signal of 2.6 (2.7) standard deviations and
a measured cross section of $\stttt=12.6^{+5.8}_{-5.2}$~fb~\cite{CMS-TOP-18-003}.

This article presents a search for \tttt production in the \SLOS\ final state using the full data set of \pp collision data at $\sqrt{s}=13$~\tev\ corresponding to 139~\ifb.
The result is then combined with that obtained in the \SSML\ final state using the same data set~\cite{Aad:2020klt}.
% End of text imported from the .//intro/intro.tex input file
 
% The next lines are included from the .//atlas/atlas.tex input file
\section{The ATLAS detector}
\label{sec:atlas}
 
\newcommand{\AtlasCoordFootnote}{
ATLAS uses a right-handed coordinate system with its origin at the nominal interaction point (IP)
in the centre of the detector and the \(z\)-axis along the beam pipe.
The \(x\)-axis points from the IP to the centre of the LHC ring,
and the \(y\)-axis points upwards.
Cylindrical coordinates \((r,\phi)\) are used in the transverse plane,
\(\phi\) being the azimuthal angle around the \(z\)-axis.
The pseudorapidity is defined in terms of the polar angle \(\theta\) as \(\eta = -\ln \tan(\theta/2)\).
Angular distance is measured in units of \(\Delta R \equiv \sqrt{(\Delta\eta)^{2} + (\Delta\phi)^{2}}\).}
 
The ATLAS experiment~\cite{PERF-2007-01} at the LHC is a multipurpose particle detector
with a forward--backward symmetric cylindrical geometry and a near \(4\pi\) coverage in
solid angle.\footnote{\AtlasCoordFootnote}
It consists of an inner tracking detector (ID) surrounded by a thin superconducting solenoid
providing a \SI{2}{\tesla} axial magnetic field, electromagnetic (EM) and hadron calorimeters, and a muon spectrometer (MS).
The inner tracking detector covers the pseudorapidity range \(|\eta| < 2.5\).
It consists of silicon pixel, silicon microstrip, and transition radiation tracking detectors.
Lead/liquid-argon (LAr) sampling calorimeters provide EM energy measurements
with high granularity.
A steel/scintillator-tile hadron calorimeter covers the central pseudorapidity range (\(|\eta| < 1.7\)).
The endcap and forward regions are instrumented with LAr calorimeters
for EM and hadronic energy measurements up to \(|\eta| = 4.9\).
The MS surrounds the calorimeters and is based on
three large air-core toroidal superconducting magnets with eight coils each.
The field integral of the toroids ranges between \num{2.0} and \SI{6.0}{\tesla\metre}
across most of the detector.
The MS includes a system of precision tracking chambers and fast detectors for triggering.
A two-level trigger system is used to select events.
The first-level trigger is implemented in hardware and uses a subset of the detector information
to accept events at a maximum rate of nearly \SI{100}{\kHz}.
This is followed by a software-based trigger that
reduces the accepted event rate to \SI{1}{\kHz}, on average,
depending on the data-taking conditions~\cite{TRIG-2016-01}.
An extensive software suite~\cite{ATL-SOFT-PUB-2021-001} is used for real and simulated data reconstruction
and analysis, for operation and in the trigger and data acquisition systems of the experiment.

% End of text imported from the .//atlas/atlas.tex input file
 
% The next lines are included from the .//selection/selection.tex input file
\section{Object and event selection}
\label{sec:selection}

Events are selected from $\sqrt{s} = \SI{13}{\TeV}$ \pp collision data collected by the ATLAS detector in the period between 2015 and 2018.
Only events for which all detector subsystems were operational are considered. The data set corresponds to an integrated luminosity of 139~\ifb~\cite{ATLAS-CONF-2019-021,Avoni:2018iuv}.
Events were required to fire single-electron or single-muon triggers,
with minimum \pT\ thresholds varying from 20 to 26 GeV depending on the lepton flavour and the data-taking period.
Triggers with minimum \pT\ thresholds include isolation requirements~\cite{TRIG-2018-05,TRIG-2018-01}.
 
Events are required to have at least one vertex reconstructed from at least two ID tracks with
transverse momenta $\pT>\SI{0.4}{\GeV}$.
The primary vertex for each event is defined as the vertex with the highest sum of \pTsq\ over all associated ID tracks~\cite{ATL-PHYS-PUB-2015-026}.
 
Object reconstruction closely follows that of Ref.~\cite{Aad:2020klt} and is briefly summarised in the following.
Electron candidates are reconstructed from energy deposits in the EM calorimeter associated with ID tracks~\cite{EGAM-2018-01}.
The pseudorapidity of the calorimeter energy cluster, $\eta_{\text{cluster}}$, must satisfy $|\eta_{\text{cluster}}|<2.47$,
excluding the transition region between the barrel and the endcap calorimeters (${|\eta_{\text{cluster}}|\not\in[1.37,1.52]}$).
Muon candidates are reconstructed by combining tracks reconstructed in both the ID and the MS~\cite{PERF-2015-10} and are required to have $|\eta|<2.5$.
Both the electron and muon candidates are required to have $\pT>$~10~\GeV.
The transverse impact parameter divided by its estimated uncertainty, ${|\dzero|/\sigma(\dzero)}$, is required to be lower
than five (three) for electron (muon) candidates. The longitudinal impact parameter $z_0$ must satisfy ${|\zzsth|<0.5}$~mm for
both lepton flavours. Electrons are required to satisfy the `Tight' likelihood-based identification criterion and
to be isolated according to the `FixedCutTight' selection criterion~\cite{EGAM-2018-01}. Muons must
satisfy the `Medium' cut-based
identification criterion and be isolated according to the `FixedCutTightTrackOnly' selection criterion~\cite{Aad:2020gmm}.
 
Jets are reconstructed from topological clusters~\cite{PERF-2014-07} of energy deposits in the calorimeters using the \antikt
algorithm~\cite{Cacciari:2008gp,Fastjet} with a radius parameter of \(R=0.4\) and calibrated as described in Ref.~\cite{PERF-2016-04}.
They are referred to as `small-$R$ jets'.
These jets are required to have $\pT>\SI{25}{\GeV}$ and $|\eta|<2.5$. To suppress the effect of additional \pp\ collisions in the
same or a nearby bunch crossing, collectively referred to as pile-up, the jets with $\pt<120$~\GeV\ and $\abseta<2.4$ must satisfy a pile-up suppression requirement based
on the output of a multivariate classifier called the jet-vertex-tagger (JVT)~\cite{PERF-2014-03}.
Events are required to pass a set of quality criteria to suppress those containing any jets arising
from non-collision sources or detector noise~\cite{ATLAS-CONF-2015-029}.
The MV2c10 multivariate algorithm~\cite{FTAG-2018-01} is used to identify jets containing $b$-hadrons.
Each jet is given a score representing the likelihood of the jet to contain a $b$-hadron.
A jet is $b$-tagged if the score passes a certain threshold, referred to as an operating point (OP).
Four OPs are defined with average expected efficiencies for $b$-jets of 60\%, 70\%, 77\% and 85\%.
A pseudo-continuous score is assigned to each jet passing these OPs,
with an integer value ranging from five for jets that pass the
60\% OP to two for jets passing only the 85\% OP.
A score of one is assigned if the jet does not pass any of the OPs.
 
The selected and calibrated small-$R$ jets are used as inputs for jet reclustering~\cite{ATLAS-CONF-2017-062}
using the \antikt algorithm with a radius parameter of $R=1.0$. These reclustered jets are referred to as `large-$R$ jets'.
The calibration corrections and uncertainties for the reclustered large-$R$ jets are inherited from the small-$R$ jets~\cite{ATLAS-CONF-2017-062}.
A trimming procedure is applied to the reclustered jets to remove all the associated small-$R$ jets that have \pT\ below 5\% of the \pT of the reclustered jet~\cite{Krohn:2009th}.
The reclustered jets are required to have $\pT>\SI{200}{\GeV}$ and $|\eta|<2.0$.
 
\newcommand{\rapidity}{
The rapidity is defined as $y=\frac{1}{2} \ln{\frac{E+p_z}{E-p_z}}$, where $E$ is the energy and $p_z$ is the longitudinal component of the momentum along the beam pipe.
}
 
A sequential overlap removal procedure defined in Ref.~\cite{Aad:2020klt} is applied to ensure that the same calorimeter
energy deposit or the same track is not associated with two or more different reconstructed objects.

The missing transverse momentum in the event, whose magnitude is denoted by \met, is defined as the negative
vector sum of the \pT\ of reconstructed and calibrated objects in the event~\cite{PERF-2016-07}.
This sum includes the momenta of the ID tracks matched to the primary vertex but not associated with any other objects.
 
The events are required to have either exactly one lepton satisfying $\pT>\SI{28}{\GeV}$ and at least seven jets (\SL channel)
or exactly two oppositely charged leptons with $\pT>\SI{28}{\GeV}$ for the leading lepton and $\pT>\SI{10}{\GeV}$ for the subleading lepton
and at least five jets (\OS channel).
In the \OS channel, the events with two same-flavour leptons must have a dilepton invariant mass
above 15~\GeV\ and outside the $Z$-boson mass window of 83--99~\GeV.
Each event is required to have at least one reconstructed lepton that matches the lepton that fired the trigger
and must contain at least two $b$-tagged jets passing the 70\% OP.
 
% End of text imported from the .//selection/selection.tex input file
 
% The next lines are included from the .//samples/samples.tex input file
\section{Monte Carlo samples}
\label{sec:samples}
 
Monte Carlo (MC) samples of simulated events were produced to model the
SM \tttt\ signal and background processes. The nominal samples are
identical to those used in Ref.~\cite{Aad:2020klt}
and are normalised using the best theory predictions available.
The \texttt{NNPDF3.0NLO}~\cite{Ball:2014uwa} PDF set was used in all
matrix element (ME) calculations if not stated otherwise.
The top-quark mass $m_\textrm{top}$ was set to 172.5~\GeV\ in all relevant samples.
 
The nominal \tttt\ signal sample was generated using
the \MGNLO~{v2.6.2}~\cite{Alwall:2014hca} generator at NLO in the strong
coupling constant \alphas with the \texttt{NNPDF3.1NLO}~\cite{Ball:2014uwa} PDF set.
The functional form of the renormalisation and factorisation scales were set
to $\muR = \muF = m_{\textrm{T}}/4$,
where $m_{\textrm{T}}$ is defined as the scalar sum of the transverse
masses $ \sqrt{m^{2}+p^{2}_{\textrm{T}}}$ of the particles generated
from the ME calculation, following Ref.~\cite{Frederix:2017wme}.
An additional sample with settings similar to the nominal was generated using a
leading-order (LO) ME, which makes more efficient use of simulation resources as
unlike the nominal it does not contain negative weight events.
The additional sample is only used for the training of the multivariate
discriminant to separate signal from background.
 
The nominal MC sample for \ttbar\ background modelling was produced with
the HVQ program~\cite{Frixione:2007nw,Frixione:2007nu} in
the \powhegbox~v2~\cite{Frixione:2007nw,Nason:2004rx,Frixione:2007vw,Alioli:2010xd}
generator at NLO in QCD in the five-flavour scheme (5FS).
The \hdamp\ parameter, which controls the \pt\ of the first additional
emission beyond the Born configuration, was set to
1.5\mtop~\cite{ATL-PHYS-PUB-2014-021,ATL-PHYS-PUB-2016-020}.
A dedicated $t\bar{t}b\bar{b}$ sample with the highest available precision,
and with the additional $b$-quarks coming from the ME,
was produced at NLO QCD accuracy in the four-flavour scheme (4FS) with the
\powhegboxRes~\cite{Jezo:2018yaf} generator and OpenLoops~\cite{Cascioli:2011va,Denner:2016kdg},
with the \texttt{NNPDF3.0nlonf4}~\cite{Ball:2014uwa} PDF set.
For this sample, a pre-release of the implementation of this process
in \powhegboxRes was provided by the authors~\cite{ttbbPowheg}. The factorisation scale was set to
$\frac{1}{2}\Sigma_{i=t,\bar{t},b,\bar{b},j}m_{\mathrm{T},i}$ (where $j$ stands for extra
partons), the renormalisation scale was set to
$\sqrt[4]{\prod_{i=t,\bar{t},b,\bar{b}} m_{\mathrm{T},i}}$,
and the \hdamp\ parameter was set
to $\frac{1}{2}\Sigma_{i=t,\bar{t},b,\bar{b}}m_{\mathrm{T},i}$ where
$m_{\mathrm{T},i}$ is the transverse mass of a given parton.

Single-top-quark production processes, i.e.\ $tW$ associated production,
t-channel and s-channel production, were modelled using
the \powhegbox~v2~\cite{Re:2010bp,Nason:2004rx,Frixione:2007vw,Alioli:2010xd,Alioli:2009je,Frederix:2012dh}
generator at NLO in QCD.
The t-channel process was generated in the 4FS with the
\texttt{NNPDF3.0NLOnf4}~\cite{Ball:2014uwa} PDF set, while for $tW$ and s-channel
processes the 5FS was used.
The diagram removal scheme~\cite{Frixione:2008yi} was employed to handle the
interference between $tW$ and \ttbar\ production~\cite{ATL-PHYS-PUB-2016-020}.
 
The \ttW\ events were generated using
the \sherpa~2.2.1~\cite{Bothmann:2019yzt} generator.
The ME was calculated for
up to one additional parton at NLO in QCD and up to two partons at LO QCD precision
using the Comix~\cite{Bothmann:2019yzt} and
OpenLoops~\cite{Cascioli:2011va,Denner:2016kdg} libraries
and merged with the \sherpa parton shower~\cite{Schumann:2007mg}
using the MEPS@NLO
prescription~\cite{Hoeche:2011fd,Hoeche:2012yf,Catani:2001cc,Hoeche:2009rj}
and a merging scale of 30 GeV.
The production of \ttH\ events was modelled using
the \powhegbox generator at NLO in QCD.
The \ttZ\ and \tWZ\ events were generated with \MGNLO~{v2.3.3} at NLO in QCD.
The other rare top-quark processes, namely \tZ, \ttWW\
and \ttt\ production, were all modelled using the \MGNLO generator at LO in QCD.
 
The production of $V+$jets events (where $V=W$ or $Z$) was simulated with
the \sherpa~2.2.1 generator using NLO-accurate MEs for
up to two partons and LO-accurate MEs for up to four partons.
Samples of diboson final states ($VV$) were also simulated with the
\sherpa~2.2.1 generator.
 
All events generated using \powhegbox\ or \MGNLO
were interfaced with \PYTHIA~{8.230}~\cite{Sjostrand:2014zea} for the parton
shower and hadronisation, using the A14 set of tuned
parameters~\cite{ATL-PHYS-PUB-2014-021} and the
\texttt{NNPDF2.3LO}~\cite{Ball:2012cx} set of PDFs.
Additional \tttt, \ttbar and single-top-quark samples were produced
by replacing \PYTHIA~{8} with \HERWIG~{7.04}~\cite{Bahr:2008pv,Bellm:2015jjp}
for parton showering and hadronisation, using the H7UE
set of tuned parameters~\cite{Bellm:2015jjp} and the
\texttt{MMHT2014LO}~\cite{Harland-Lang:2014zoa} PDF set.
These samples are used to evaluate uncertainties due to the choice
of parton shower and hadronisation model.
 
To assess the uncertainty due to the choice of generator, the \ttbar, single-top-quark, \ttH and \ttW
samples produced with the nominal generator set-ups are
compared with alternative samples generated with \MGNLO
for the calculation of the hard-scattering, interfaced with \PYTHIA~{8}.
An alternative sample of \ttZ events was generated with NLO MEs
using \SHERPA~2.2.1. All alternative samples were generated using the same PDF in the ME
as in the nominal sample.

The effects of pile-up were modelled by overlaying minimum-bias events,
simulated using the soft QCD processes of \PYTHIA~{8.186}
with the A3 tune~\cite{ATL-PHYS-PUB-2016-017}, on events from hard processes.
For all samples of simulated events, except those generated using \SHERPA, the \EVTGEN~1.2.0
program~\cite{Lange:2001uf} was used to describe the decays of bottom and charm hadrons.

The non-\ttbar samples were processed through the simulation~\cite{SOFT-2010-01} of the
ATLAS detector geometry and response using \textsc{Geant4}~\cite{Agostinelli:2002hh}.
A fast simulation (AtlFast 2) of the ATLAS detector relying on parameterised showers in
the calorimeter~\cite{ATL-SOFT-PUB-2018-002} was used for the \tttt and \ttbar samples.
All samples were reconstructed using the same software used for collider data.
Corrections were applied to the simulated events so that the physics objects' selection
efficiencies, energy scales and energy resolutions match those determined from data
control samples.

% End of text imported from the .//samples/samples.tex input file
 
% The next lines are included from the .//strategy/strategy.tex input file
\section{Analysis strategy}
\label{sec:strategy}

In the \SLOS final state, the parton-level \tttt topology is characterised by four $b$-quarks resulting from
the decay of the four top quarks and by either six or four other quarks coming from the hadronic $W$-boson decays.
Consequently, the signature of the events from \tttt production in the ATLAS detector features a high number
of jets (\Nj), with 10 (8) jets in the \SL (\OS) channel, among which four contain a $b$-hadron.
This phase-space region is contaminated by a large number of background events coming almost exclusively
from \ttjets\ production.
After the preselection described in Section~\ref{sec:selection},
the contribution from other backgrounds is below 8\% and arises mainly from
\ttW, \ttZ, \ttH, single top-quark or $W/Z$ boson production in association with jets.
Diboson production and rare processes including \tZq, \tWZ, \ttWW and \ttt contribute less than 1\% of all events.
All small non-\ttbar\ backgrounds are estimated from simulations.
The contribution from the background arising from misreconstructed or non-prompt leptons is negligible in the \SL channel.
This has been verified by the agreement between data and MC simulation using different isolation requirements on the leptons.
In the \OS channel, this background is expected to be dominated by \ttbar\ events with one prompt lepton.
The contribution is found to be below 2\% using \ttbar\ MC simulations and is included in the \ttbar\ background.
 
The nominal \ttbar\ sample was generated at NLO as described in Section~\ref{sec:samples},
with up to one additional parton from the ME calculation.
All additional jets, including the heavy-flavour ones, are produced in the parton shower.
The background predictions in high \Nj regions were found to be unreliable.
Reference~\cite{TOPQ-2018-15} demonstrates clear mismodelling of the number of jets not from the \ttbar\ decay
and the scalar sum of the transverse momenta of the reconstructed hadronically and leptonically decaying top quarks, $H_\mathrm{T}^{\ttbar}$.
In addition, the rate of \ttbar\ production in association with $b$-jets was
observed to be underestimated by the current MC simulations~\cite{TOPQ-2017-12,CMS-TOP-18-011}.
Therefore, the following strategies were developed in this analysis to obtain a reliable estimate of the \ttjets\ background by reweighting the \ttbar\ MC samples using data.

Firstly,
the \ttjets\ MC events are classified according to the flavour of the particle jets
that are not from the decay of the \ttbar system,
using the same procedure as described in Ref.~\cite{HIGG-2013-27}.
The particle jets are reconstructed from the simulated stable particles
using the anti-\kt algorithm with a radius parameter $R=0.4$,
and are required to have $\pt>15$~\GeV\ and $\abseta<2.5$.
Events are labelled as \ttbjets\ if at least one particle jet is matched within $\Delta R < 0.4$ to
a $b$-hadron with $\pt>5$~\GeV\ that did not originate from the decay of a top quark.
Similarly, events not categorised as \ttbjets, and where at least one particle jet is matched
within $\Delta R<0.4$ to a $c$-hadron with $\pt>5$~\GeV\ that did not originate from the decay
of a top quark, are labelled as \ttcjets.
The remaining events are labelled as \ttljets.

The selected events are categorised into different regions according to
the lepton and jet multiplicities and different $b$-tagging requirements.
The details of the event categorisation can be found in Section~\ref{sec:categorisation}.
These regions vary in their signal-to-background ratio and the flavour composition of \ttjets.
Corrections to the normalisation of \ttljets, \ttcjets and \ttbjets and their kinematic distributions
are derived from regions with low signal contamination using data,
as described in Section~\ref{sec:background}.
These corrections are used to mitigate the mismodelling of \ttjets\ MC events in regions with a high signal-to-background ratio,
where a binned profile likelihood fit is performed to extract the \tttt\ signal strength.
In the regions most sensitive to \tttt production, a multivariate discriminant is used to
improve the sensitivity by distinguishing between signal and background events, as discussed in Section~\ref{sec:mva}.
Large-$R$ jets are used as proxies for hadronically decaying top quarks with high \pt,
improving the discrimination between \tttt\ signal and \ttjets\ background.

The different \ttjets\ components after the corrections are further adjusted and constrained in the profile likelihood fit.
The \ttbjets events are divided into further categories to define their systematic uncertainties.
Events with one particle jet matched to a single $b$-hadron are labelled as \ttb,
those with one particle jet matched to at least two $b$-hadrons are labelled as \ttB,
and those with two particle jets, each matched to one $b$-hadron are labeled as \ttbb.
The rest of the \ttbjets events are labelled as \ttbbb.
Section~\ref{sec:systematics} documents all systematic uncertainties considered in this analysis.
Details of the scheme used for the \ttjets\ modelling systematic uncertainties are presented in Section~\ref{subsec:ttbarUnc}.

% End of text imported from the .//strategy/strategy.tex input file
 
% The next lines are included from the .//categorisation/categorisation.tex input file
\section{Event categorisation}
\label{sec:categorisation}
 
Selected events in both channels are categorised into several regions
as illustrated in Figure~\ref{fig:categorisation}.
The resulting regions are used for different purposes.
Signal-depleted and signal-enriched regions are defined as control regions and signal regions, respectively.
They are used in the profile likelihood fit to extract the signal cross section
and to constrain the overall background model using the assigned systematic uncertainties.
The \ttjets kinematic reweighting regions are used to extract correction factors for the \ttjets MC prediction.
The validation regions are defined in order to check that the background prediction is
able to describe data in regions with background composition similar to
that in the signal regions.
 
To take advantage of the high jet multiplicity in the \tttt\ signal, events in each lepton channel
are first categorised according to their number of jets, from 7 (5) jets to at least 10 (8) jets in the \SL
(\OS) channel.
These events are then further categorised according to the $b$-tagging requirements summarised in
Table~\ref{tab:btagging}.
The requirements were chosen to provide good separation
between the different flavour components of the associated jets in the \ttjets background.
The 2b, 4b and $\ge$5b regions are defined by respectively requiring the presence of 2, 4 and at least
5 jets $b$-tagged at the 70\% OP.
The $\ge$5b regions in the \SL channel help constrain the modelling of \ttbbb.
In the \OS channel, the 4b and $\ge$5b regions are merged into $\ge$4b regions
because fewer events are expected.
The events with 3 jets $b$-tagged at the 70\% OP are further assigned to the 3bL, 3bH and 3bV
regions using requirements on the number of jets $b$-tagged at the 60\% and 85\% OPs.
The 3bL (3bH) regions are defined to have relatively lower (higher) purity of MC `truth' $b$-jets amongst
the three jets tagged at the 70\% OP.\footnote{MC `truth' $b$-jets are defined as jets that are matched to $b$-hadrons.}
As a result, the 3bH regions contain a larger fraction of \ttb and \ttB events,
whereas the 3bL regions are more populated by \ttcjets and \ttljets,
where the third jet $b$-tagged at the 70\% OP is a mis-tagged $c$-jet or light-flavour jet.
The 3bV regions are defined to be orthogonal to 3bL and 3bH regions and are used to validate the background modelling
in events enriched in \ttbjets.
 
\begin{table}[!hbt]
\begin{center}
\caption{
Summary of the $b$-tagging requirements for the event categorisation.
Events in each category must satisfy all requirements listed in columns.
$N_b^{60\%}$, $N_b^{70\%}$ and $N_b^{85\%}$ are defined as the numbers of $b$-tagged jets
obtained using $b$-tagging operating points with average
expected efficiencies of 60\%, 70\% and 85\%, respectively.
The 3bL (3bH) requirement selects events with lower (higher) purity of
MC `truth' $b$-jets amongst the three jets tagged at the 70\% OP.
The 3bV requirement is used to define the validation regions.
The symbol `-' indicates that no requirement is applied.
\newline
}
\begin{tabular}{m{6em}m{5em}m{5em}m{5em}}
\hline
\\[-1em]
Name & $N_b^{60\%}$ &  $N_b^{70\%}$ & $N_b^{85\%}$ \\
\\[-1em]
\hline
2b             & -       & = 2     & - \\
3bL            & $\le$ 2 & = 3     & - \\
3bH            & = 3     & = 3     & = 3 \\
3bV            & = 3     & = 3     & $\ge$ 4 \\
$\ge$4b (\OS) & -       & $\ge$ 4 & - \\
4b (\SL)        & -       & = 4     & - \\
$\ge$5b (\SL)   & -       & $\ge$ 5 & - \\
\hline
\end{tabular}
\label{tab:btagging}
\end{center}
\end{table}
 
\begin{figure}[!h]
\centering
\includegraphics[width=0.49\textwidth]{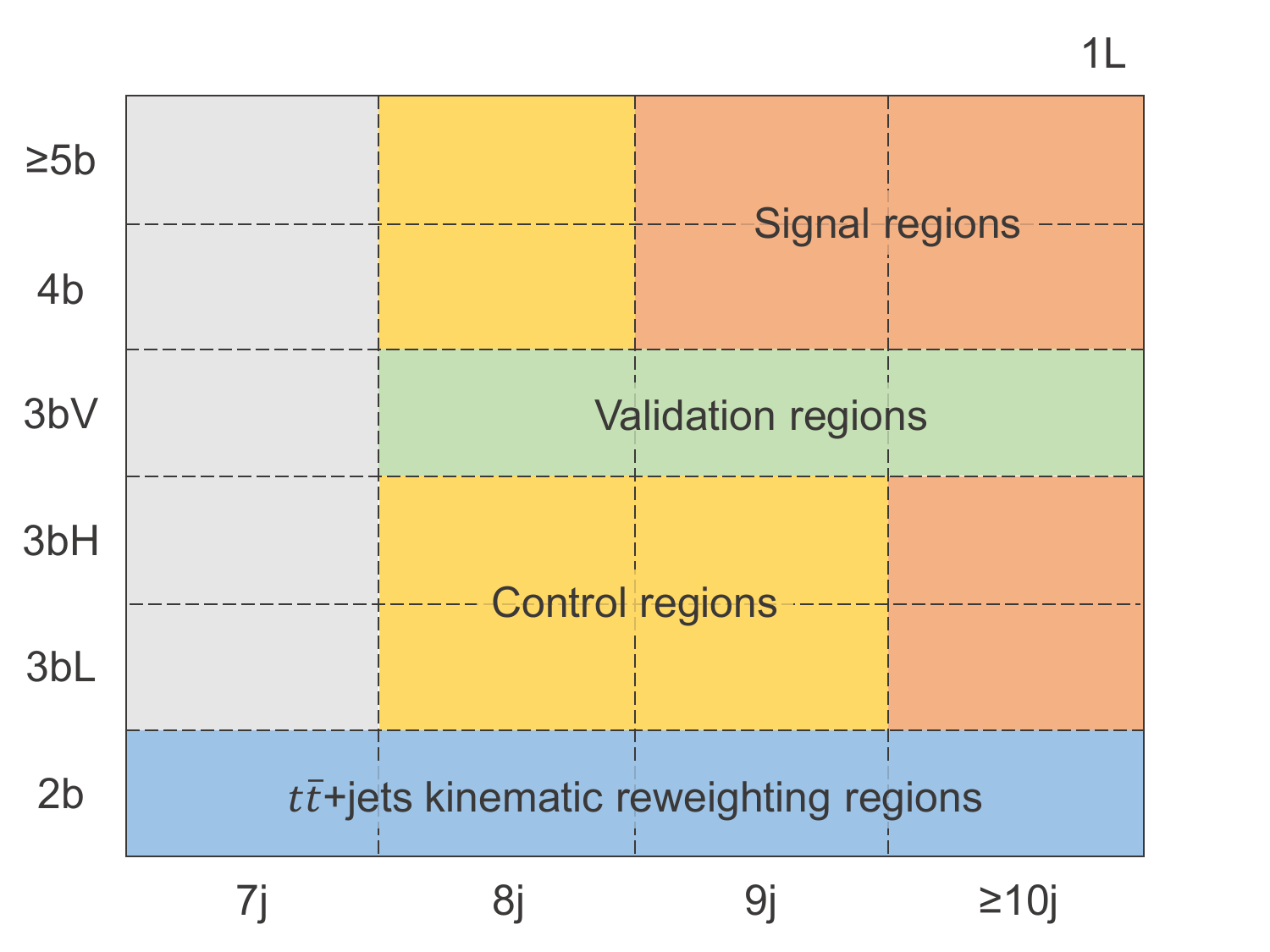}
\includegraphics[width=0.49\textwidth]{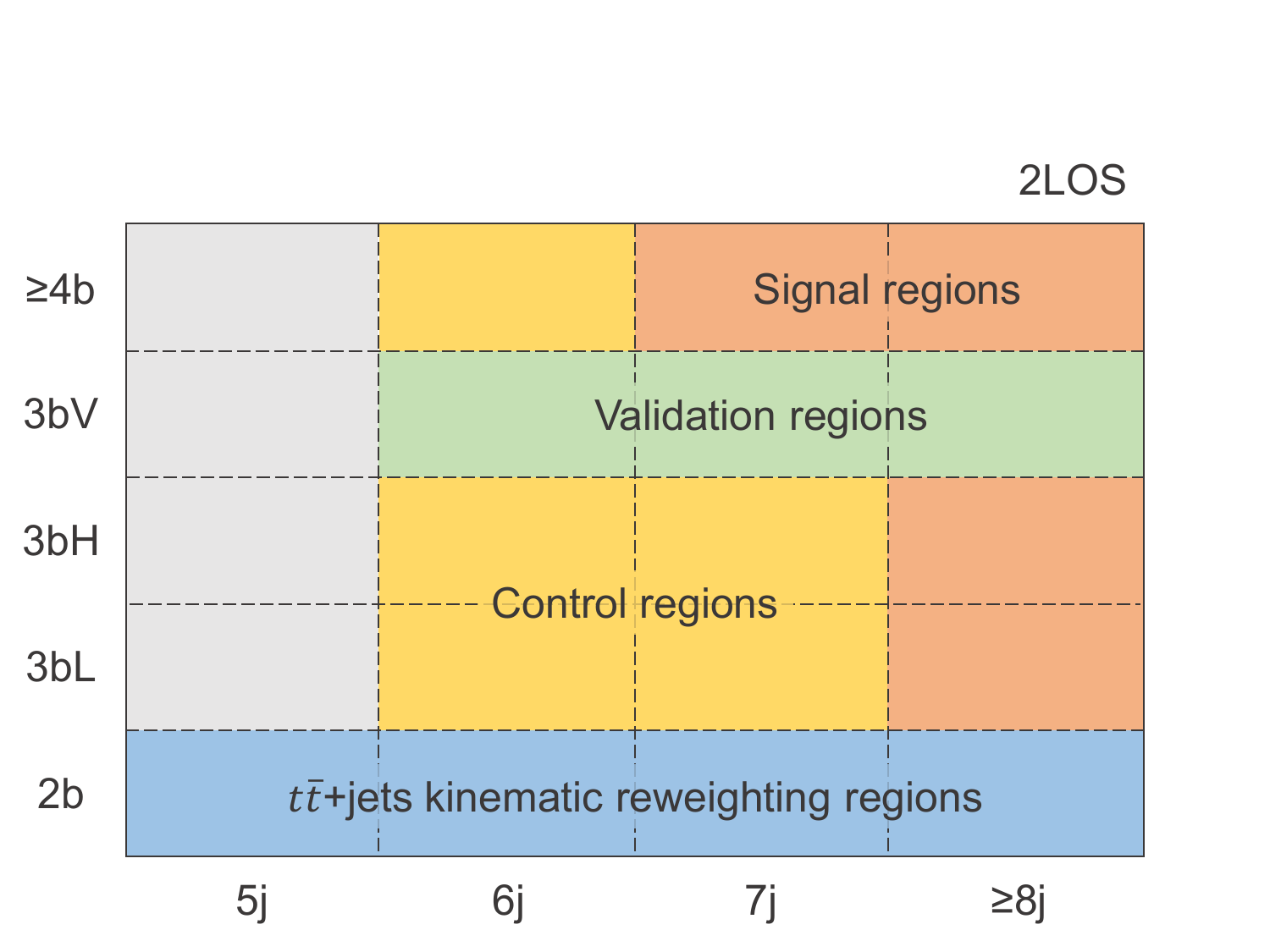}
\caption{Schematic view of the event categories used to select analysis regions
(signal, control, validation and tt+jets reweighting regions)
in the \SL channel (left) and \OS channel (right).
The axes represent the jet multiplicity and $b$-tagging requirements defined in Table~\ref{tab:btagging}.
The 3bL (3bH) $b$-tagging requirement selects events with lower (higher) purity of MC
`truth' $b$-jets amongst the three jets tagged at the 70\% OP.
The 3bV $b$-tagging requirement is used to define the validation regions.
The regions in grey are not used in the analysis.
}
\label{fig:categorisation}
\end{figure}
 
A total of 21 regions are used in the profile likelihood fit, with 12 regions in the \SL channel and
9 regions in the \OS channel.
They are defined by considering the regions with at least 8 (6) jets in the \SL (\OS) channel and
satisfying the 3bL, 3bH or $\ge$4b requirements.
Among these regions, the ones that have at least 10 (8) jets or have 9 (7) jets and satisfy the $\ge$4b
requirement in the \SL (\OS) channel are defined as the signal regions.
The rest of the fitted regions are defined as the control regions.
A total of 6 validation regions are also defined by considering the regions with at least 8 (6)
jets in the \SL (\OS) channel and satisfying the 3bV requirement.
The validation regions are not used in the fit and hence do not contribute to the signal extraction.
The largest signal contamination in the validation regions is expected to be 4.4\% in the ($\geq$8j, 3bV) region in the \OS channel.
A test found that including these regions in the fit would increase the sensitivity to the signal by 5\%.
However, this small gain in sensitivity is relinquished to ensure reliable background modelling.
The \ttjets kinematic reweighting regions are defined by considering all the regions that satisfy the
2b requirement.
 
Figure~\ref{fig:composition} shows the background composition in each signal, control and validation region,
as expected from the nominal \ttjets MC simulation
after applying the corrections described in Section~\ref{sec:background}.
The contribution from the signal is included, assuming the cross section predicted by the SM.
The control regions have a signal contamination of no more than 1\%.
The largest signal-to-background ratio evaluated from the inclusive yields
in the signal regions is 6.1\%.
 
\begin{figure}[!h]
\centering
\includegraphics[width=0.49\textwidth]{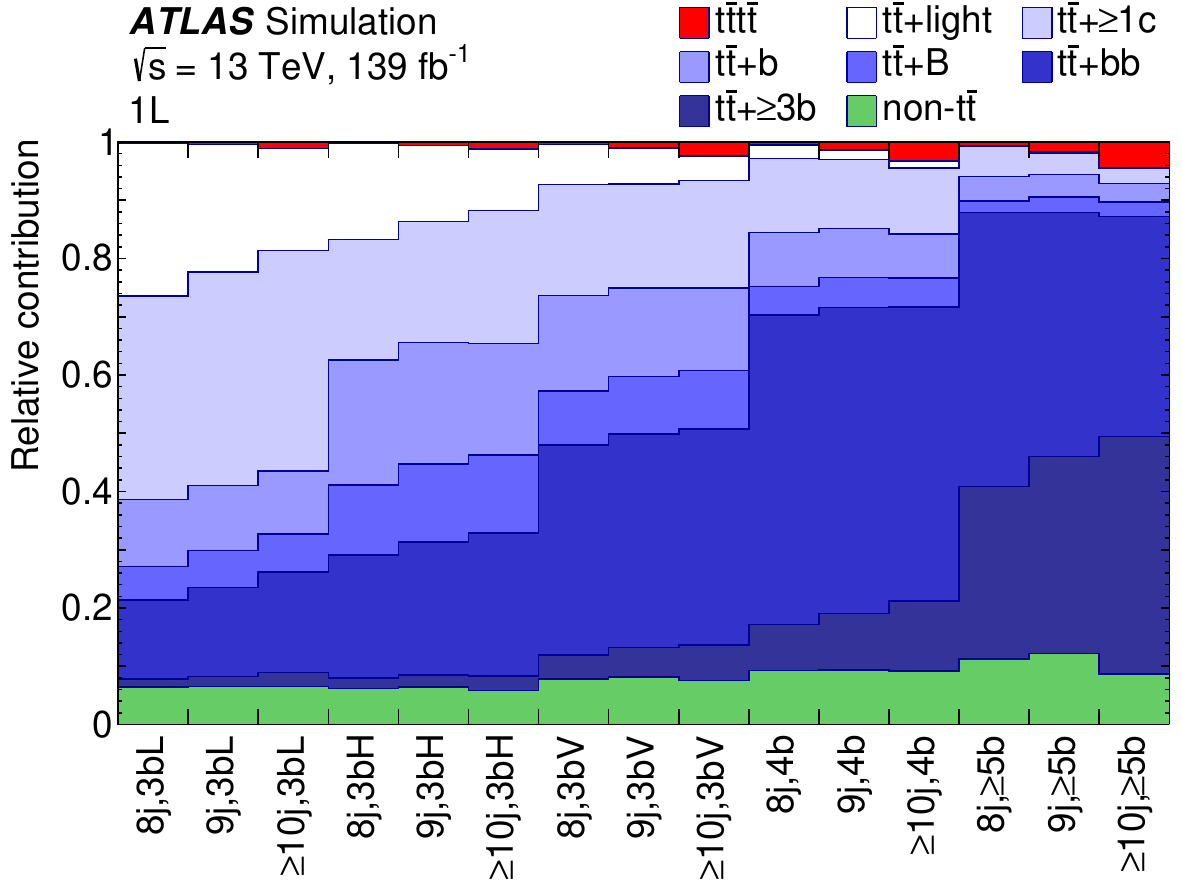}
\includegraphics[width=0.49\textwidth]{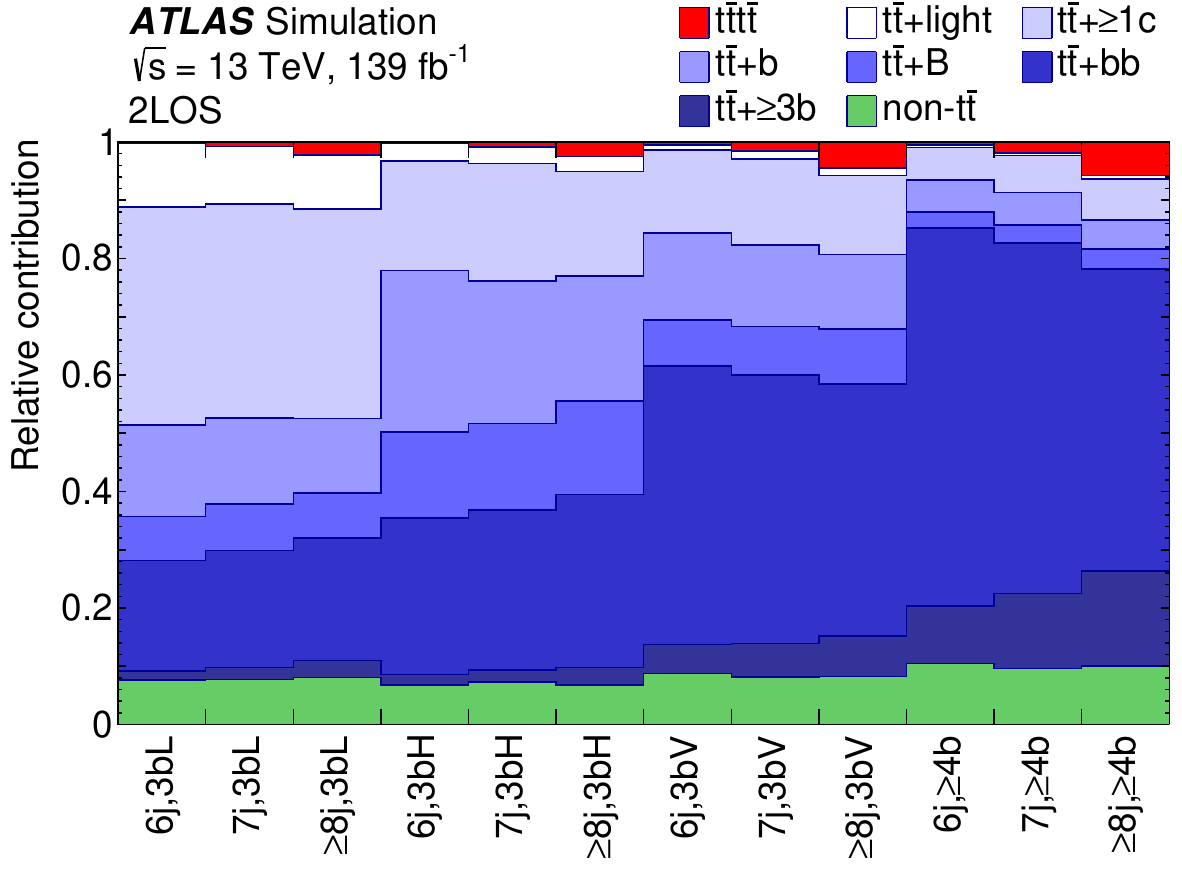}
\caption{Relative contribution from the signal and backgrounds in all signal, control and validation regions
in the \SL channel (left) and \OS channel (right).
The 3bL (3bH) $b$-tagging requirement selects events with lower (higher) purity of
MC `truth' $b$-jets amongst the three jets tagged at the 70\% OP.
The 3bV $b$-tagging requirement is used to define the validation regions.
For the \ttjets background, the fraction is shown for each component with the finer classification.
The \tttt signal is normalised to the SM cross-section prediction.
}
\label{fig:composition}
\end{figure}
 
% End of text imported from the .//categorisation/categorisation.tex input file
 
% The next lines are included from the .//background/background.tex input file
\section{Modelling of \ttjets\ background}
\label{sec:background}

This section describes the corrections applied to both the nominal \ttjets prediction
and to the associated systematic uncertainties.
The corrections applied to the \ttjets\ MC sample include a rescaling of
the \ttjets\ flavour components and a sequential kinematic reweighting.
Any possible residual mismodelling is accounted for by the systematic
uncertainties in the profile likelihood fit used for the
extraction of the signal strength.
 
\subsection{\ttjets flavour rescaling}
 
The \ttjets\ flavour rescaling adjusts the overall yields of the \ttljets, \ttcjets and \ttbjets categories.
The rescaling factors are derived from a dedicated profile likelihood fit to data
using the event yields in the regions defined by various $b$-tagging requirements.
Events with $\geq8$j in the \SL channel and $\geq6$j in the \OS channel are assigned to 2b, 3bL, 3bH and $\geq$4b regions,
using the same criteria as defined in Table~\ref{tab:btagging}.
The flavour rescaling fit exploits the different \ttjets\ flavour fractions in the eight fitted regions.
These regions have the same \Nj requirements as the regions used in the profile likelihood fit to extract the signal.
The largest signal-to-background ratio in these regions is 2.5\%, estimated from MC simulation prior to the fit.
Systematic uncertainties due to the tagging efficiency of $b$-jets and the mis-tag rate of $c$-jets and light-flavour jets
are considered as nuisance parameters.
The measured rescaling factors for \ttljets, \ttcjets and \ttbjets are $0.99 \pm 0.05$, $1.58 \pm 0.18$ and $1.33 \pm 0.06$, respectively,
where the quoted uncertainties are from the statistical uncertainty of the data and from uncertainties in the $b$-tagging calibration.
 
The measured rescaling factors are applied to the \ttjets\ MC sample entering the profile likelihood fit to extract the signal;
however, their measured uncertainties are not used in the fit.
Instead, large global normalisation uncertainties are assigned to \ttHF\ events, as discussed in Section~\ref{subsec:ttbarUnc}.
However, \ttjets\ modelling uncertainties due to the choice of generator and scale
also affect the flavour composition of the radiated jets.
To reduce the correlation among these modelling uncertainties and the dedicated \ttHF\ normalisation uncertainties,
the \ttjets\ flavour fraction in all modelling systematic variations is rescaled
to match the nominal prediction after the flavour rescaling within the acceptance of the analysis,
i.e.\ events in the \SL\ (\OS) channel with $\Nj\geq8~(6)$ and $N_b^{70\%}\geq2$.
The \ttbjets\ events in the $t\bar{t}b\bar{b}$ 4FS sample are rescaled to have the same yield as the nominal \ttbjets events within the acceptance.
Such treatment ensures that the overall production rate of \ttHF\ events is controlled by the global normalisation uncertainties,
whilst the modelling systematic uncertainties affect only the relative changes in the yield across different analysis regions
and the kinematic distributions within each region.

\subsection{Sequential kinematic reweighting}
 
Following the flavour rescaling, a sequential reweighting is used to mitigate the kinematic mismodelling observed in \ttjets MC simulation.
The reweighting corrects the distributions of \Nj, the number of large-$R$ jets (\Nrcj),
the scalar sum of all jet and lepton \pt\ in the event (\htall), and the average \dR\ between any two jets (\dRavgjj).
These variables are related to the overall jet activity in the events and are observed to be mismodelled, especially the \Nj and \htall spectra.
These variables capture the most representative global kinematics of the events,
as well as kinematic properties of the individual jets such as their \pt\ and angular distributions.
 
The \ttjets\ events in $\geq$3b regions are reweighted so the overall MC prediction matches the data in the 2b regions.
In this procedure it is assumed that the radiation modelling deficiency in the parton shower
is independent of the flavour of the radiated jets.
Systematic variations of the \ttjets\ modelling cover possible deviations from this assumption.
 
The reweighting is performed in three steps, separately for the \SL and \OS channels.
In the first step, the two-dimensional distribution of the numbers of jets and large-$R$ jets, (\Nj, \Nrcj), is reweighted.
The \Nj spectrum is corrected up to $\geq$ 13 ($\geq$ 11) jets, while the \Nrcj spectrum is corrected up to $\geq$ 2 ($\geq$ 1) large-$R$ jets in the \SL (\OS) channel.
The reweighting factors are derived for each (\Nj, \Nrcj) bin.
The second step aims to correct for the mismodelling in the \htall\ spectrum.
The reweighting is performed on the \textit{reduced \htall} variable,
defined as \htred = $\htall - (\Nj - N_{\mathrm{min}})\times90$~\GeV,
where $N_{\mathrm{min}}=7~(5)$ in the \SL (\OS) channel.
The value of 90~\GeV\ corresponds to the average \pt\ of each additional jet,
estimated as the difference between the most probable values of \htall\ in consecutive \Nj\ bins.
The term subtracted from \htall\ therefore removes the dependence of this variable on \Nj
and minimises the correlation between the first and the second reweighting steps.
The reweighting factors are therefore derived in an inclusive region with $\Nj \geq 7~(5)$ in the \SL\ (\OS) channel,
but separately for each \Nrcj multiplicity bin, from 0 to $\geq$ 2 ($\geq$ 1) large-$R$ jets.
The events with 7 (5) jets in the \SL (\OS) channel are included in this step
to increase the statistical precision of the reweighting factors in the highest \Nrcj multiplicity bin.
These events are not used elsewhere in the analysis.
To further reduce statistical fluctuations in the binned reweighting factors of \htred,
a three-parameter exponential function, $f(\htred) = a + b \cdot \exp{(c \cdot \htred)}$,
is used to fit the factors in each \Nrcj multiplicity bin.
The final step corrects for the residual mismodelling in the angular distributions
by reweighting the binned distribution of \dRavgjj, the average \dR between any two jets.
This is performed in each (\Nj, \Nrcj) bin, up to $\geq$ 10 ($\geq$ 8) jets and $\geq$ 2 ($\geq$ 1) large-$R$ jets
in the \SL\ (\OS) channel.
 
Uncertainties in the derived reweighting factors are propagated as systematic uncertainties in the \ttjets\ background.
The uncertainty sources include the limited numbers of data and MC events and the cross-section uncertainties of the non-\ttbar processes.
An additional uncertainty is considered for the \htred\ reweighting step by replacing the exponential function with a reciprocal function.
A more detailed discussion of these uncertainties is presented in Section~\ref{subsec:ttbarUnc}.
 
The reweighting factors are also derived for each systematic uncertainty affecting the \ttjets\ prediction,
such that in the 2b regions all systematic variations match the nominal prediction and thus the data.
After applying the reweighting, the overall systematic uncertainty in the $\geq$3b regions is reduced.

\subsection{Effects of the corrections to \ttjets}
 
Figure~\ref{fig:correction_effect} illustrates the combined effect of the flavour rescaling and the sequential reweighting.
The \Nj and \htall distributions for events with $\geq$ 8 jets and $\geq3$ $b$-jets in the \SL\ channel are shown as examples.
Significant mismodelling is present in both distributions before the corrections.
A large improvement in the level of agreement between data and MC simulation is achieved
for both the overall normalisation and the kinematic shape.
The flavour rescaling factors increase the overall normalisation of \ttcjets and \ttbjets,
which dominate the deficiency of MC events in the $\geq$3b regions.
The sequential reweighting increases the jet multiplicity
while decreasing the total energy of the event and the average distance between the jets predicted by the \ttjets\ simulation.
The total uncertainty in the MC prediction is also reduced after the
sequential reweighting is applied coherently
to the \ttjets\ systematic variations.
The level of agreement after the reweighting of \dRavgjj is equally good,
though the effect of the correction is smaller.
 
\begin{figure}[!p]
\centering
\vspace{-2.0cm}
\includegraphics[width=0.45\textwidth]{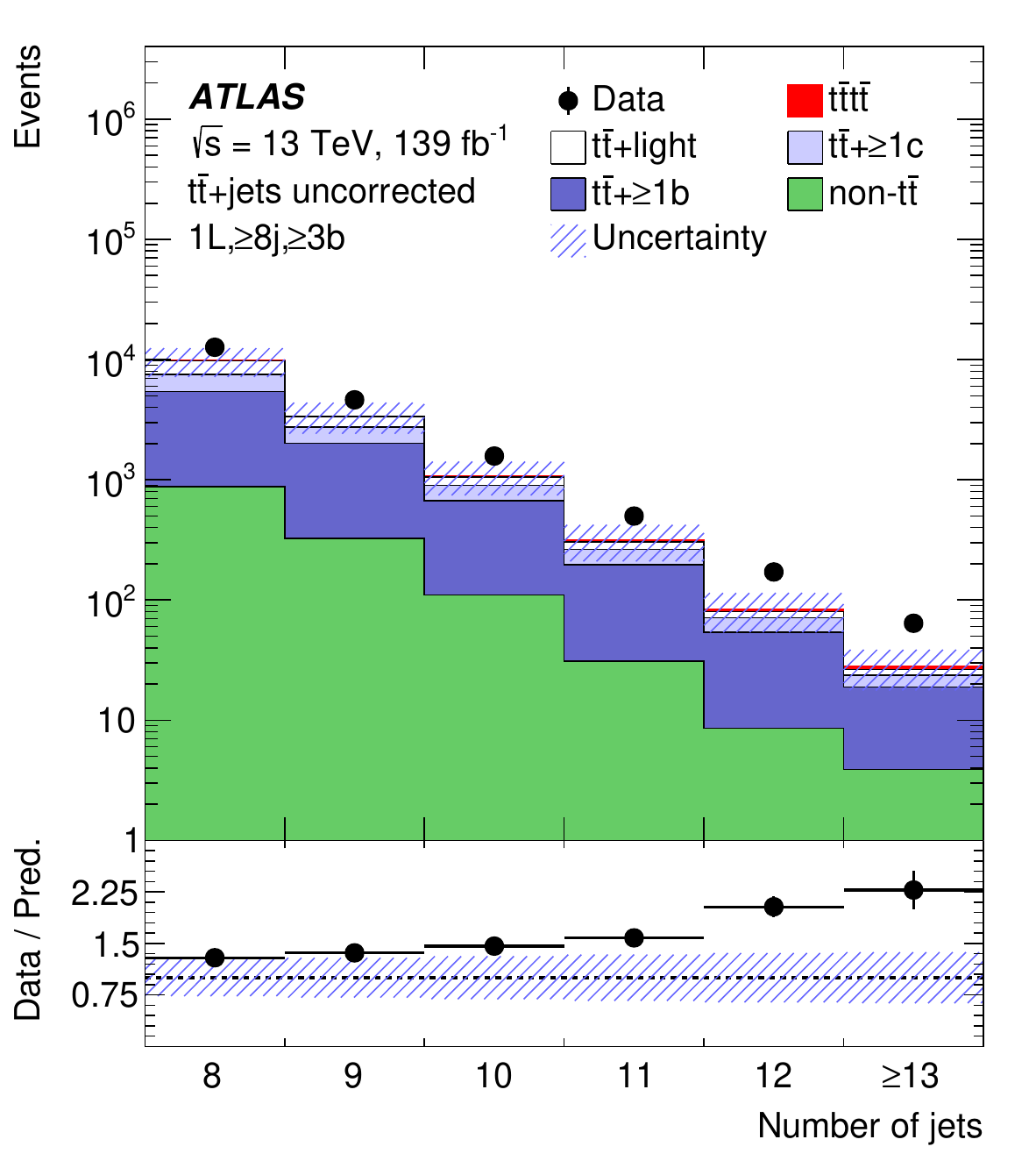}
\includegraphics[width=0.45\textwidth]{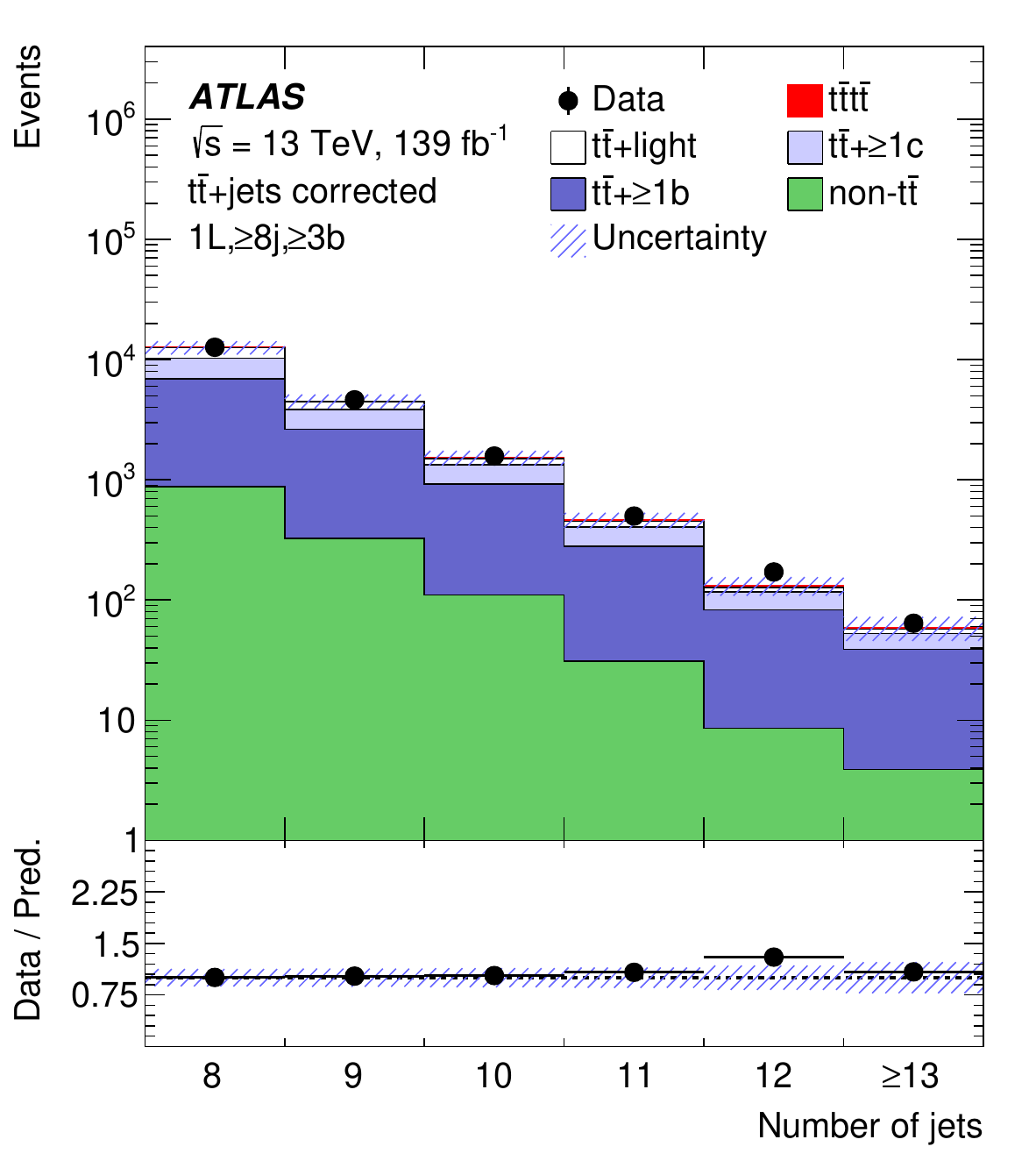}\\
\includegraphics[width=0.45\textwidth]{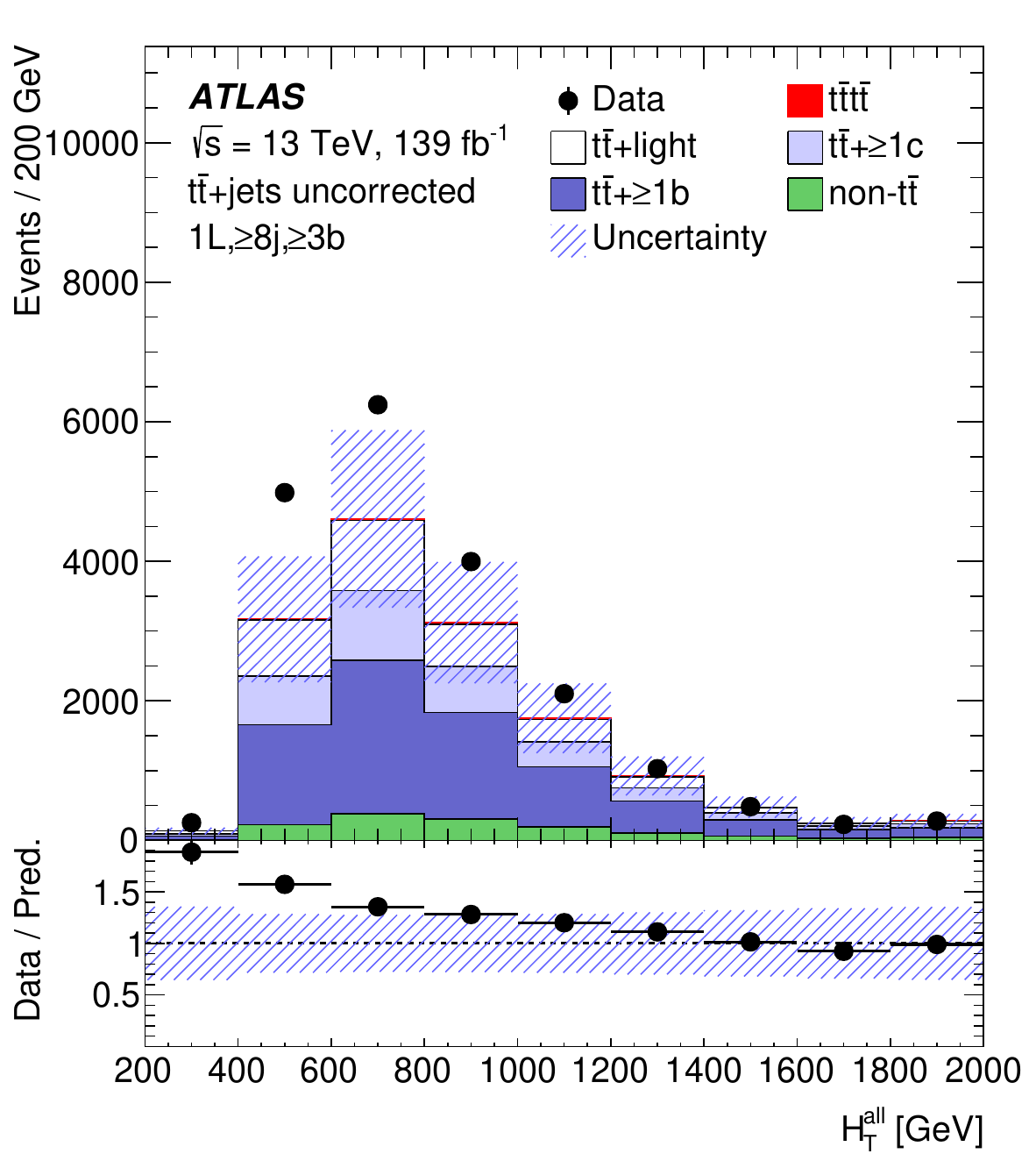}
\includegraphics[width=0.45\textwidth]{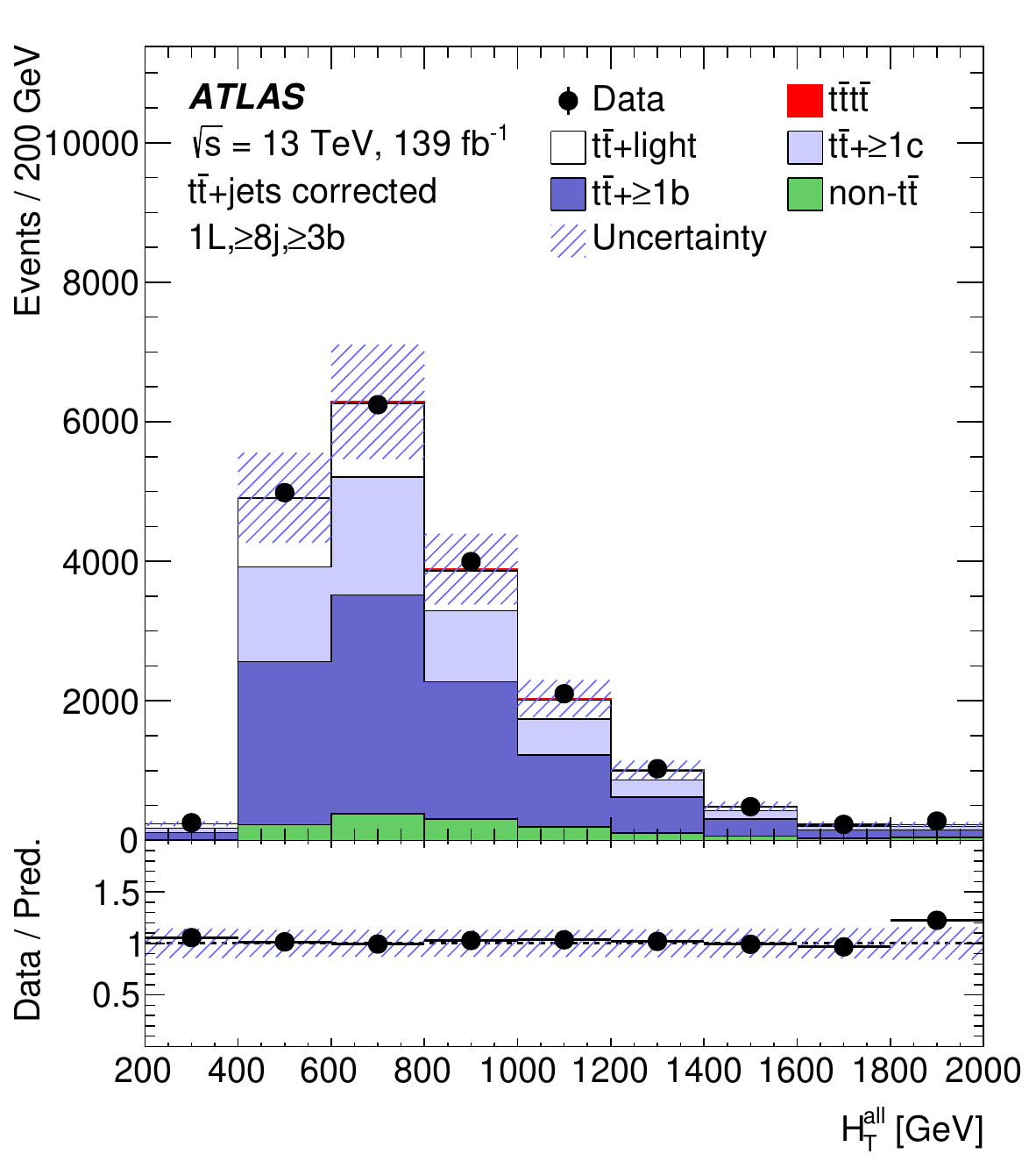}
\caption{The \Nj and \htall distributions in the region with $\geq$ 8 jets and $\geq$ 3 $b$-jets in the \SL channel
before (left) and after (right) the flavour rescaling and the sequential kinematic reweighting.
The band includes the total pre-fit uncertainty of the MC prediction.
The ratio of the data to the total MC expectation is shown in the lower panel.
The last bin in all distributions includes the overflow.
}
\label{fig:correction_effect}
\end{figure}
 
% End of text imported from the .//background/background.tex input file
 
% The next lines are included from the .//mva/mva.tex input file
\section{Signal discrimination}
\label{sec:mva}

A multivariate analysis is performed in the signal regions to discriminate
\tttt\ signal from the large background.
It uses boosted decision trees (BDTs)~\cite{TMVA} combining several input observables
to build an output score maximising the separation between signal and the total
background.
The LO \tttt\ simulated signal sample is used in the training,
while the BDT performance is evaluated on the NLO \tttt sample.
For the distributions of kinematic variables used by the multivariate discriminant,
good agreement was observed when comparing them in LO and NLO simulation.
The background sample is composed of the simulated \ttjets\ events after the data-driven corrections are applied
and the non-\ttbar backgrounds as predicted by the simulation.
 
The BDTs are trained in six regions using fourteen variables. The six regions are defined
by events with at least three $b$-tagged jets and 8 (6), 9 (7) and at least 10 (8) jets in the
\SL (\OS) channel.
The BDTs use global event variables,
the kinematic properties of reconstructed objects and pairs of objects,
jet $b$-tagging information,
the multiplicity and substructure of large-$R$ jets, and the \met.
The most powerful discriminating variable in all regions
is the sum of the pseudo-continuous $b$-tagging score
over the six jets with the highest score in the event.
The definition of the pseudo-continuous $b$-tagging score can be found
in Section~\ref{sec:selection}.
In the highest \Nj regions in the two channels, where an inclusive requirement on
the number of jets is applied, the jet count is an important discriminating variable.
The minimum distance $\Delta R$ among all pairs of $b$-tagged jets also provides
good discrimination in the majority of regions since the spatial separation
of $b$-jets in the \tttt\ signal is larger on average than in the main \ttbjets\
background. Other variables include \htall calculated using all reconstructed jets
and leptons in the event, centrality $\sum_{i}{\pt}_{i}/\sum_{i}E_{i}$, where
the sum runs over all reconstructed jets and leptons in the event, leading jet
\pt, minimum $\Delta R$ among all pairs of $b$-tagged jets and leptons,
average $\Delta R$ between all pairs of jets, the invariant mass of the
triplet of jets that has the smallest $\Delta R$,\footnote{The $\Delta R$
of a triplet of jets is defined as
$\Delta R_{ijk} = \sqrt{\Delta R^2_{ij}+\Delta R^2_{ik}+\Delta R^2_{jk}}$, where
$i$, $j$, $k$ are the indices of the three jets.} the \met, and the transverse mass
of the $W$ boson, $m_\mathrm{T}(\ell, \met)$, in the single-lepton channel.\footnote{The transverse mass is defined as
$\sqrt{2\pt\met(1-\cos\Delta\phi)}$, where $\Delta\phi$ is the azimuthal angle
between the lepton and \met.}
Additional variables are related to large-$R$ jets: the number of large-$R$ jets with a
mass above 100~\GeV, the sum of the first \kt\ splitting scale
$d_{12}$ of all large-$R$ jets,\footnote{The \kt\ splitting scale $d_{ij}$ is defined
as the recombination distance between the jet constituents from
a \kt\ algorithm with radius parameter $R$: $d_{ij}=\min({\pt}_i^2,{\pt}_j^2)\times\Delta R_{ij}^2/R^2$.} and the
sum of the second \kt\ splitting scale $d_{23}$ of all large-$R$ jets.
 
The modelling of the input variables was checked in the control and validation regions
before and after the fit by propagating the fitted parameters obtained from the fit
in the control and signal regions to the validation regions.
Figure~\ref{fig:btag} shows the modelling of the sum of the pseudo-continuous
$b$-tagging score in each lepton channel before performing the fit.
The distribution in each lepton channel is shown for an inclusive selection of at least
three $b$-tagged jets and at least 9 (7) jets in the \SL (\OS) channel.
These regions include events from the validation, control and signal regions where the
BDT score distribution is used.
Taking into account all uncertainties, no significant discrepancy between data
and the predicted background was found.
 
\begin{figure}[!h]
\centering
\includegraphics[width=0.45\textwidth]{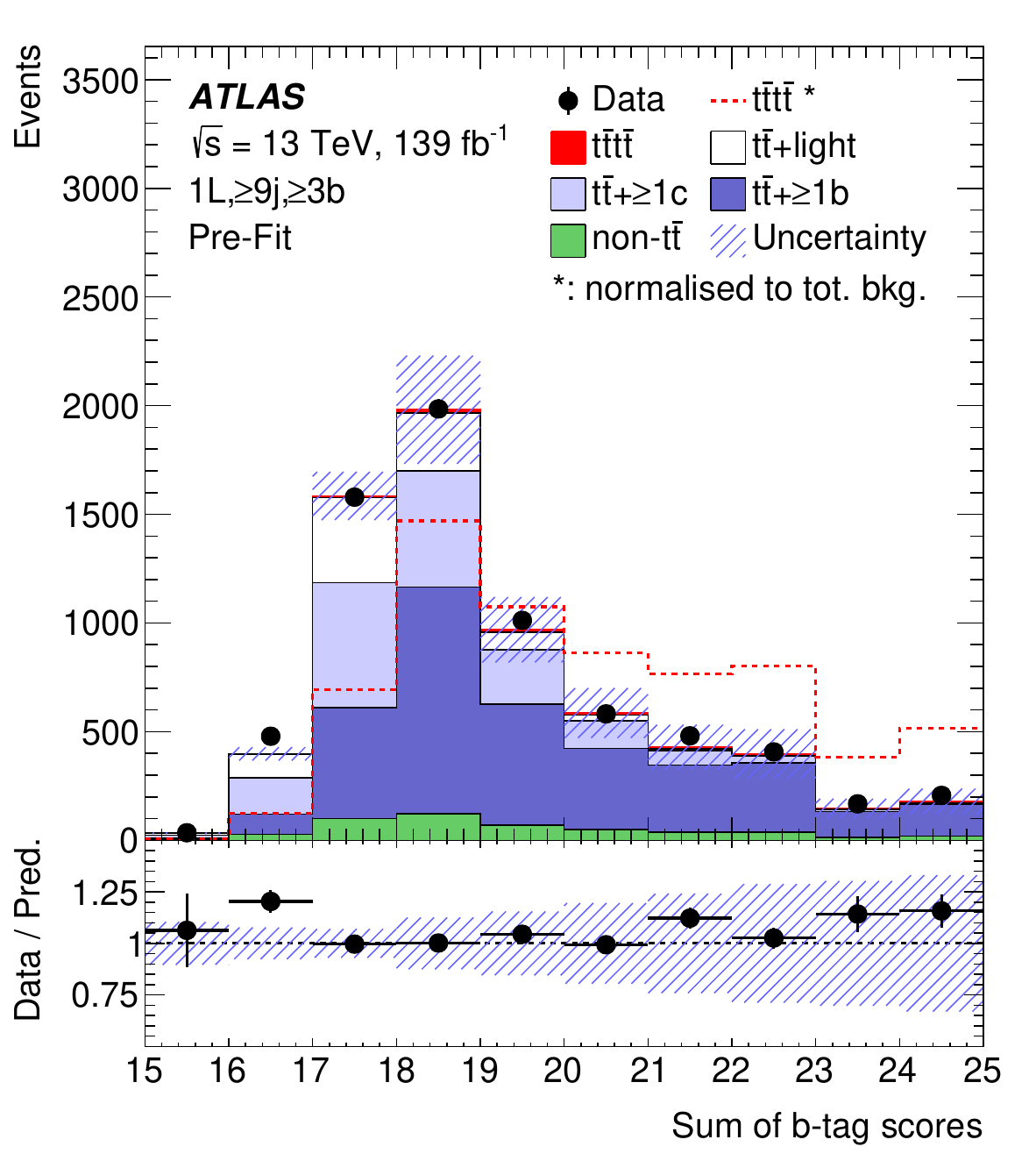}
\includegraphics[width=0.45\textwidth]{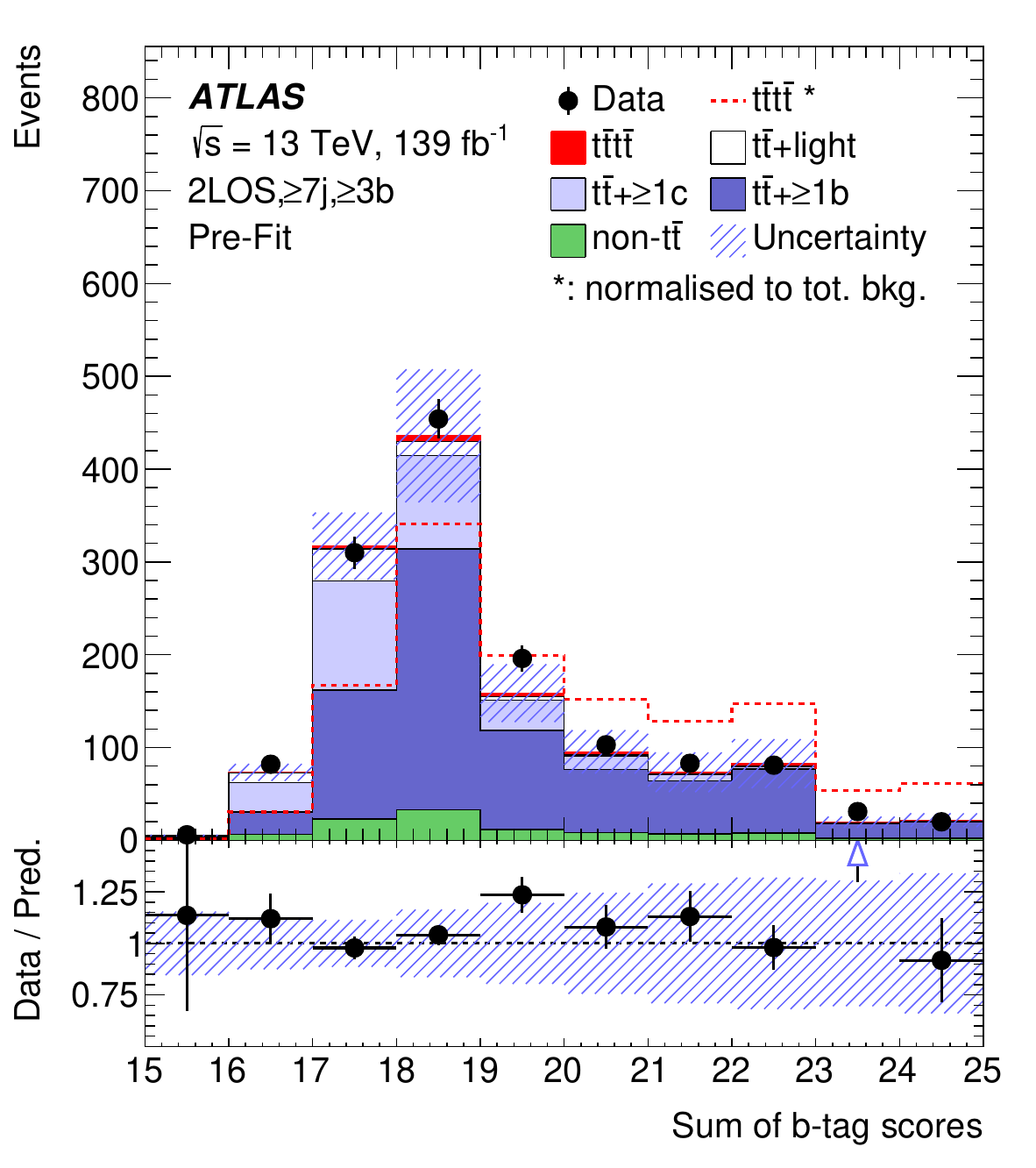}
\caption{
Pre-fit comparison between data and prediction for the distributions of the sum of the pseudo-continuous
$b$-tagging score over the six jets with the highest score in the event for the \SL channel (left)
and the \OS channel (right)
in the regions with $\geq$ 3 $b$-jets and $\geq$ 9 (7) jets in the \SL (\OS) channel.
The \ttjets background is corrected using data.
The band includes the total uncertainty of the pre-fit computation.
The dashed red line shows the signal distribution normalised to the background yield.
The ratio of the data to the total pre-fit expectation is shown in the lower panel.
The last bin contains overflow events.
}
\label{fig:btag}
\end{figure}

% End of text imported from the .//mva/mva.tex input file
 
% The next lines are included from the .//systematics/systematics.tex input file
 
\section{Systematic uncertainties}
\label{sec:systematics}
 
Various sources of systematic uncertainty affect the estimated signal and
background rates, including those related to the luminosity, the
identification and reconstruction of the physics objects, referred to as
experimental uncertainties, and the modelling
of the signal and background processes. In the following, a brief
description of the sources of systematic uncertainty is provided.
Particular emphasis is placed on the uncertainties related to the \ttbar
background prediction, which has the largest impact on the
sensitivity of the measurement.
 
% The next lines are included from the .//systematics/experimental.tex input file
\subsection{Experimental uncertainties}
\label{subsec:expUnc}
 
The experimental uncertainties are the same as in Ref.~\cite{Aad:2020klt}
and are briefly summarised here.
The uncertainty in the combined 2015--2018 integrated luminosity is $1.7\%$~\cite{ATLAS-CONF-2019-021},
obtained using the LUCID-2 detector~\cite{Avoni:2018iuv} for the primary luminosity measurements.
An uncertainty in the pile-up simulation is derived from the
uncertainty in the scale factors used to
adjust the MC pile-up to the data pile-up profile.
Uncertainties in the modelling of electrons and muons in MC simulation
arise from their momentum scale calibration and resolution, as well as
trigger, reconstruction, identification, and isolation efficiencies.
Uncertainties in the modelling of jets are primarily related to their energy scale (JES)
and resolution (JER). The JES uncertainty is decomposed into a set of 30 uncorrelated
components referred to as eigenvectors (EV),
with contributions from pile-up, jet flavour composition and
single-particle response~\cite{ATL-PHYS-PUB-2015-015}. The JER uncertainty is
represented by nine components~\cite{PERF-2011-03}.
An uncertainty in the efficiency to pass the JVT requirement for suppressing pile-up jets
is also considered~\cite{PERF-2014-03}.
The $b$-tagging efficiencies and mis-tagging rates are measured in data using the
same methods as those described in
Refs.~\cite{FTAG-2018-01,ATLAS-CONF-2018-001,ATLAS-CONF-2018-006}.
The uncertainties in the $b$-tagging calibration are determined separately for
$b$-jets, $c$-jets and light-flavour jets. They account for differences
between data and simulation, depending on \pt for $b$- and $c$-jets, and
on \pt and $\eta$ for light-flavour jets, and are divided into 85 components
(45 for $b$-jets, 20 for $c$-jets and 20 for light-flavour jets).
The \met uncertainty due to a possible miscalibration of its soft-track component
is derived from data--MC comparisons of the \pT balance between the hard and
soft \met components~\cite{PERF-2016-07}.
% End of text imported from the .//systematics/experimental.tex input file
% The next lines are included from the .//systematics/signal.tex input file
\subsection{Signal modelling uncertainties}
\label{subsec:signalUnc}
 
Various sources of modelling uncertainty are considered for the \tttt signal.
An uncertainty related to the missing higher-order QCD corrections is estimated by
varying the renormalisation (\muR) and the factorisation (\muF) scales simultaneously by
factors of $2.0$ and $0.5$ relative to the nominal value.
The effect of the PDF variations on the signal MC prediction was evaluated
following the PDF4LHC prescription~\cite{Botje:2011sn} and found to be negligible.
The uncertainty due to the choice of parton shower and hadronisation model is
estimated by comparing the nominal \tttt MC sample with the alternative that
uses \HERWIG~{7} to simulate the PS, as described in Section~\ref{sec:samples}.
 
% End of text imported from the .//systematics/signal.tex input file
% The next lines are included from the .//systematics/ttbarModel.tex input file
\subsection{Uncertainties in the \ttjets background}
\label{subsec:ttbarUnc}
 
The systematic uncertainties affecting the \ttjets\ background modelling are
summarised in Table~\ref{tbl:ttbarSys_description}. They are applied to each flavour
component of the \ttbar background separately, i.e.\ treated as uncorrelated, to
account for the variation in flavour composition of the regions included in the fit
and for possible differences in the modelling of the
\ttbjets, \ttcjets, and \ttljets processes.
 
An uncertainty of $50\%$ in the
normalisation of \ttcjets events as well as of the different subcategories of the \ttbjets
events (\ttb, \ttbb, \ttB, and \ttbbb) is applied~\cite{TOPQ-2017-12}.
The uncertainties due to the choice of generator and PS model used to simulate
the inclusive \ttbar sample are evaluated by comparing the nominal \ttbar sample with
the alternative \ttbar samples, detailed in Section~\ref{sec:samples}.
They are split into shape and migration components, with the former affecting
only the shape of the distributions in each fit region and the latter changing the
yields in different regions. These uncertainties are applied to each of the
\ttbjets subcategories as detailed in Table~\ref{tbl:ttbarSys_description}.

Uncertainties due to missing higher-order QCD corrections are estimated by separately
varying the renormalisation and the factorisation scales by factors of $2.0$ and $0.5$
in the nominal \ttbar sample.
Additionally, uncertainties in the amounts of initial- and final-state radiation (ISR and FSR)
from the PS are assessed by respectively varying the corresponding
parameter of the A14 PS tune\footnote{The A14 tune's variations
of $\alpha_\mathrm{S}^\mathrm{ISR}(m_Z)$ are usually referred to as Var3c, and correspond
to $\alpha_\mathrm{S}^\mathrm{ISR}(m_Z)=0.127^{+0.013}_{-0.012}$~\cite{ATL-PHYS-PUB-2014-021}.}
and by varying the FSR renormalisation scale by factors of $2.0$ and $0.5$.
An additional uncertainty in the main background \ttbjets process is evaluated by comparing
its modelling by the nominal sample, where $b$-quarks are produced in the parton shower,
with the prediction by the \powhegboxRes generator which calculates the \ttbb process in
the matrix element at NLO in QCD with massive $b$-quarks
in the 4FS.
 
\begin{table}[htbp]
{\scriptsize
\begin{center}
\caption{Summary of the sources of systematic uncertainty for the \ttjets\ modelling.
The last column of the table lists the uncorrelated components of each systematic uncertainty source.
All systematic uncertainty sources are treated as uncorrelated across the \ttjets components.
For generator and PS choices, each \ttjets\ component is further decomposed into a shape component and a migration component.
The number of uncorrelated components for each physics source is shown in parentheses.
}
\begin{tabular}{l c c}
\toprule
Uncertainty source     & Description                                                 & Components (number) \\ \midrule
\ttbjets normalisation & $\pm 50\%$                                                  & \ttb, \ttbb, \ttB, \ttbbb (4) \\
\ttcjets normalisation & $\pm 50\%$                                                  & \ttcjets (1) \\ \midrule
Generator choice       & \powheg vs \MGNLO                                          & (\ttljets, \ttcjets, \ttb, \ttbb, \ttB, \ttbbb) \\
& & $\otimes$ (shape, migration) (12) \\
PS choice              & \PYTHIA~{8} vs \HERWIG~{7}                                    & (\ttljets, \ttcjets, \ttb, \ttbb, \ttB, \ttbbb) \\
& & $\otimes$ (shape, migration) (12) \\
Renormalisation scale  & Varying \muR in \powheg                       & \ttljets, \ttcjets, \ttbjets (3) \\
Factorisation scale    & Varying \muF in \powheg                       & \ttljets, \ttcjets, \ttbjets (3) \\
ISR                    & Varying $\alpha_{\textrm{S}}^{\textrm{ISR}}$(PS) in \PYTHIA~{8} & \ttljets, \ttcjets, \ttbjets (3) \\
FSR                    & Varying \muF(PS) in \PYTHIA~{8}                   & \ttljets, \ttcjets, \ttbjets (3) \\
5FS vs 4FS            & \powhegboxRes (4FS) vs \powhegbox (5FS)                    & \ttb, \ttbb, \ttB, \ttbbb (4) \\
\hline
\end{tabular}
\label{tbl:ttbarSys_description}
\end{center}
}
\end{table}
 
A group of uncertainties arise from the kinematic reweighting
procedure applied to correct for the mismodelling of the \ttjets\ background.
The uncertainties due to the limited numbers of data and MC events
in each of the reweighted bins of the (\Nj, \Nrcj) and \dRavgjj\ variables
result in 120 systematic uncertainties.
Their effect on \htall\ and the BDT discriminants is less than 1\% in any bin of the corresponding distribution.

The uncertainty in the \htred\ reweighting is evaluated using
a 68\% confidence interval contour of the nominal fit to the exponential function,
resulting in five systematic uncertainties corresponding to the five fitted
functions in each of the \Nrcj\ multiplicity bins in the two channels.
Five additional systematic uncertainties are obtained
by replacing the exponential function with a three-parameter reciprocal function,
$f(\htred) = a + b / c^{\htred}$.
The difference compared to the exponential function is smaller than 20\%
in any fitted regions, with a larger deviation at high or low \htred.
The net effect of the systematic uncertainties in the \htred\ reweighting
is below 5\% across the \htall\ and BDT distributions.
 
The reweighting factors depend on the contribution from the non-\ttbar\
processes in the 2b regions which is subtracted before deriving the reweighting.
The cross sections of these processes are varied by $\pm$50\% simultaneously,
and the reweighting factors are rederived. The resulting difference is taken as
an additional systematic uncertainty, ranging from 3\% to 6\%, in the reweighting factors.

% End of text imported from the .//systematics/ttbarModel.tex input file
% The next lines are included from the .//systematics/nonttbar.tex input file
\subsection{Modelling uncertainties in non-\ttbar backgrounds}
\label{subsec:nonttbarUnc}
 
An uncertainty of 5.5\% in the total cross section of the three
single-top-quark production modes is
included~\cite{Kidonakis:2010tc,Kidonakis:2010ux,Kidonakis:2011wy}.
Similar to the \ttbar background, uncertainties associated with the PS and hadronisation model and the generator choice
are evaluated by comparing the nominal \powheg+~\PYTHIA~{8} sample for each process with
alternative samples produced with \powheg+~\HERWIG~{7} and \MGNLO+~\PYTHIA~{8}.
The uncertainty due to the treatment of the interference between \tW and \ttbar
production at NLO is evaluated by comparing the
nominal $tW$ sample produced using the diagram removal scheme with an alternative
sample produced with the same generator but using the diagram subtraction scheme~\cite{Frixione:2008yi}.
 
Modelling uncertainties in the \ttW, \ttZ, and \ttH processes are evaluated in a
similar way. Uncertainties of $60\%, 15\%,$ and $20\%$ are applied
to the \ttW, \ttZ, and \ttH cross sections, respectively~\cite{deFlorian:2016spz,HIGG-2018-57,Aad:2020klt}.
The PDF uncertainty in the \ttW, \ttZ and \ttH processes is negligible
compared to the uncertainty applied to their cross section.
The uncertainty due to missing higher-order QCD corrections is determined
independently for each of these processes
by varying simultaneously the renormalisation and factorisation scales
by factors of $2.0$ and $0.5$ relative to the nominal value.
The uncertainty due to the choice of generator and PS
is assessed by comparing the nominal simulation of each of these processes with the
alternative one described in Section~\ref{sec:samples}.
Additional uncertainties of 10\%, 20\% and 30\% for events with
respectively 9 (7), 10 (8) and at least 11 (9) jets are applied to account for
the production of additional jets in these processes in the \SL (\OS) channel.
These uncertainty values are based on the level of mismodelling of the
\ttjets \Nj distribution observed in the 2b region. They are also consistent with the
modelling uncertainties evaluated in the \SSML\ final state for these backgrounds using a
full set of MC modelling uncertainties~\cite{Aad:2020klt}.
 
An uncertainty of 60\% is assumed for the \Vjets production cross section.
It is estimated by adding a 24\% uncertainty in quadrature for each additional jet
based on a
comparison among different algorithms for merging LO matrix elements
and parton showers~\cite{Alwall:2008epjc}.
 
For each of the other small background processes a conservative cross-section uncertainty of 50\% is applied.

% End of text imported from the .//systematics/nonttbar.tex input file

% End of text imported from the .//systematics/systematics.tex input file
 
% The next lines are included from the .//results/results.tex input file
\section{Results}
\label{sec:results}

The analysis uses a binned profile likelihood fit to extract the \tttt\ signal strength
with systematic uncertainties parameterised as constrained nuisance parameters.
The regions used in the fit are described in Section~\ref{sec:categorisation}.
In all control regions, the binned \htall\ distribution is used as the input to the profile likelihood fit,
except for the (8j, $\ge$5b) region in the \SL channel where the event count is used
due to the small number of events and the low signal-to-background ratio in this region.
To improve the sensitivity to the \tttt signal,
the distribution of the BDT score described in Section~\ref{sec:mva} is used in the signal regions.
Each bin in these regions is represented by a Poisson probability term
for the observed data, with the expectation value provided by the MC prediction with the corrections to \ttjets.
The expected yield depends on the signal strength, $\mu$,
defined as the ratio of the measured \tttt\ production cross section to that predicted by the SM, $\sigma_{\tttt}/\sigma^{\text{SM}}_{\tttt}$,
and the systematic uncertainties, $\pmb{\theta}$.
A binned likelihood function, $\mathcal{L}(\mu, \pmb{\theta})$,
is constructed as the product of the Poisson probability terms of all fitted bins.
Systematic uncertainties are taken into account in the likelihood function,
each as a nuisance parameter constrained by a Gaussian prior.
The uncertainties due to the limited number of generated MC events,
also referred to as the MC statistical uncertainties, are implemented as
additional nuisance parameters with Poisson priors, one for each fitted bin.
The MC prediction is fitted to the data by adjusting $\mu$ and $\pmb{\theta}$ simultaneously to maximise the likelihood function.
The statistical inferences of the fit results are derived using the likelihood ratio,
$\lambda(\mu) = \mathcal{L}(\mu, \hat{\hat{\pmb{\theta}}})/\mathcal{L}(\hat\mu, \hat{\pmb{\theta}})$,
where $\hat\mu$ and $\hat{\pmb{\theta}}$ are the values of $\mu$ and $\pmb{\theta}$ that maximise the likelihood function
and $\hat{\hat{\pmb{\theta}}}$ represents the values that maximise the likelihood function given a specific $\mu$~\cite{Cowan:2010js}.
The uncertainty in the best-fit $\mu$ value is determined by finding the $\mu$ values that correspond to $-2\ln\lambda(\mu) = 1$.
The test statistic $q_0 = -2\ln\lambda(0)$ is used to check the compatibility of the background-only hypothesis ($\mu=0$) with the observed data,
and consequently the significance of the fitted signal.
The statistical analysis is performed using the \textsc{RooStat} framework~\cite{RooFit,moneta2011roostats}.

Figure~\ref{fig:summary} shows the observed event yield compared with the prediction in each control
and signal region in each lepton channel before the fit (`pre-fit') and after the fit (`post-fit').
The level of agreement between data and the prediction is improved post-fit due to the adjustment
of the nuisance parameters.
The overall uncertainty in the prediction is also significantly reduced as a result of the constraints from the data
and the correlations among the nuisance parameters.
Figures~\ref{fig:sr_postfit_1l} and \ref{fig:sr_postfit_2l} show the distribution of the BDT score
in the signal regions in both lepton channels.
Good agreement is observed between data and the post-fit prediction in all fitted regions.
The goodness of fit was evaluated using a likelihood-ratio test,
comparing the likelihood value from the nominal fit with the one obtained from a saturated model,
built with one free-floating normalisation factor for each fitted bin~\cite{Cousins2013GeneralizationOC}.
The probability that the fit model is compatible with the observed data is 59\%.
The post-fit background modelling is further checked in validation regions that are not used in the fit.
The definition of these regions is described in Section~\ref{sec:categorisation}.
Figure~\ref{fig:vr_postfit} shows the distribution of the BDT score in four of those regions.

\begin{figure}[!htb]
\centering
\includegraphics[width=0.49\textwidth]{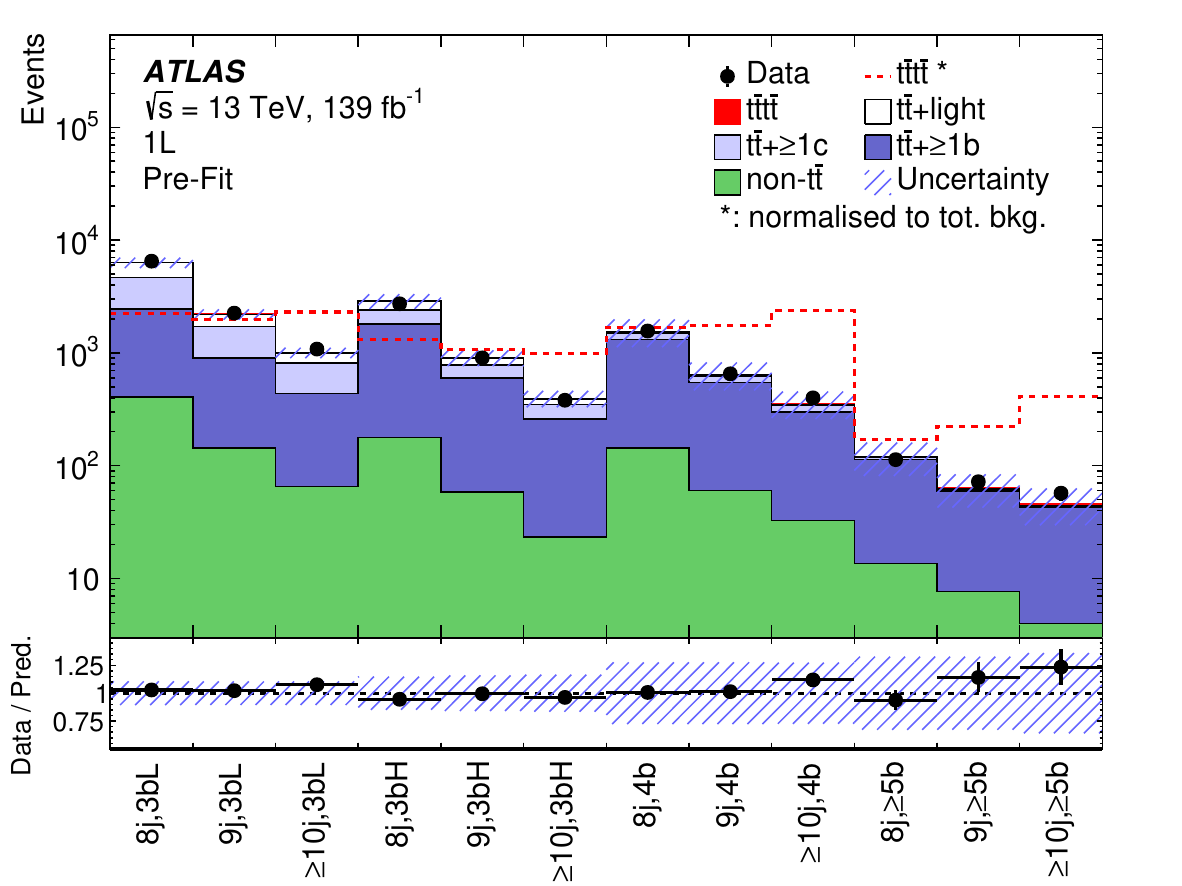}
\includegraphics[width=0.49\textwidth]{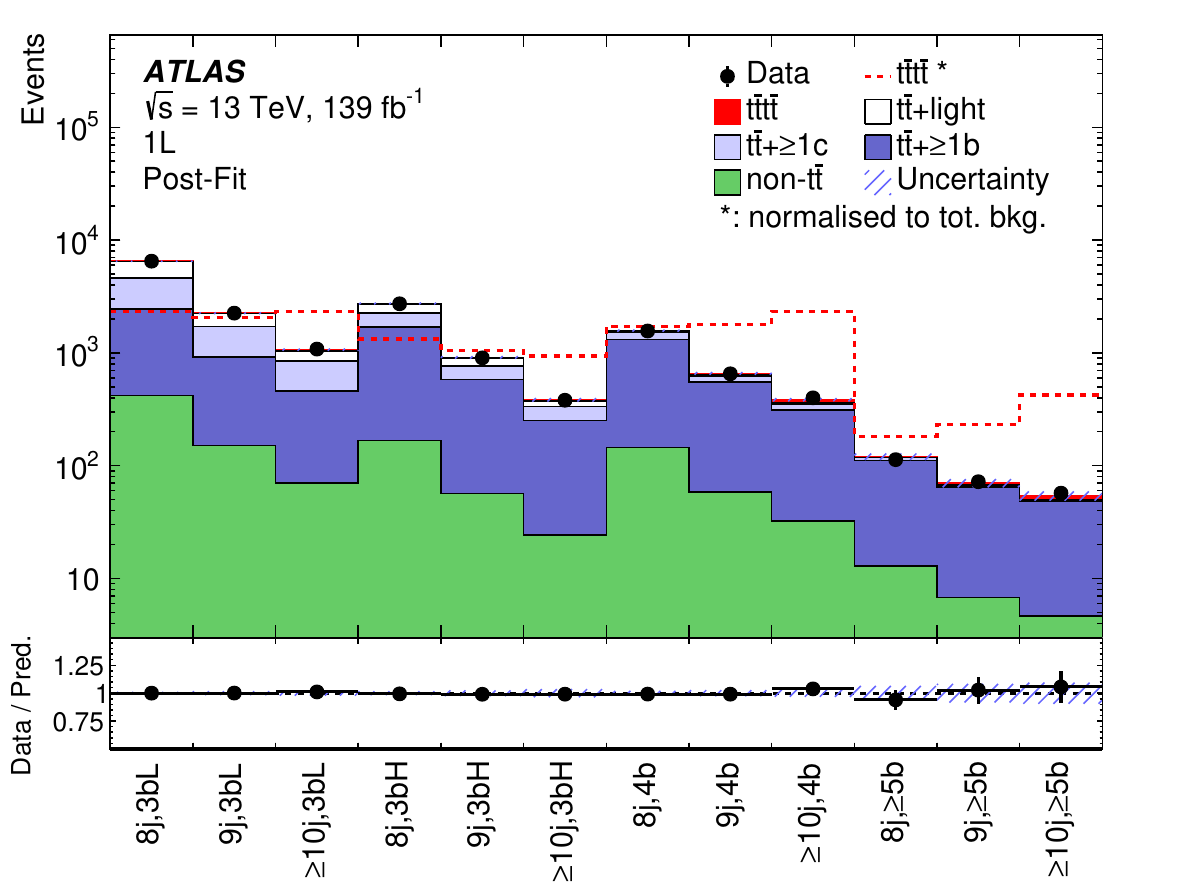}
\includegraphics[width=0.49\textwidth]{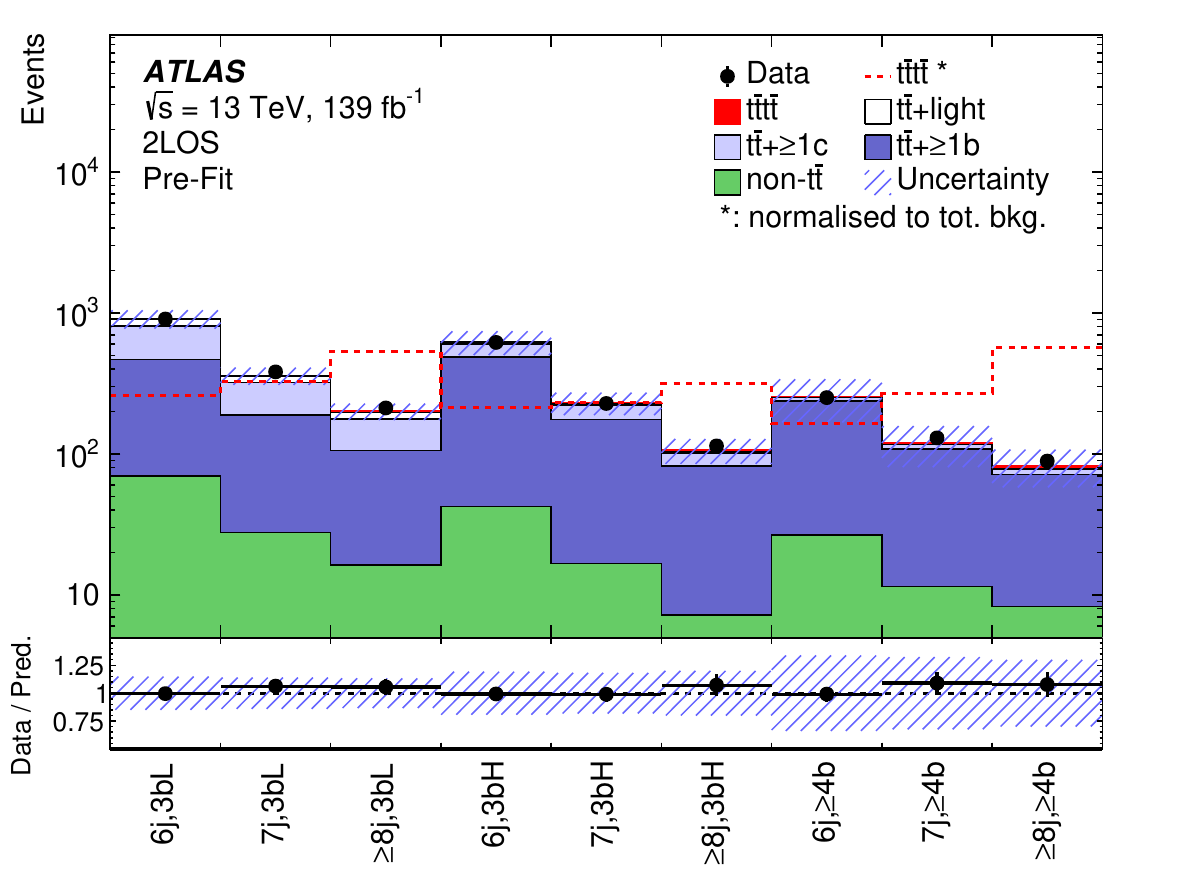}
\includegraphics[width=0.49\textwidth]{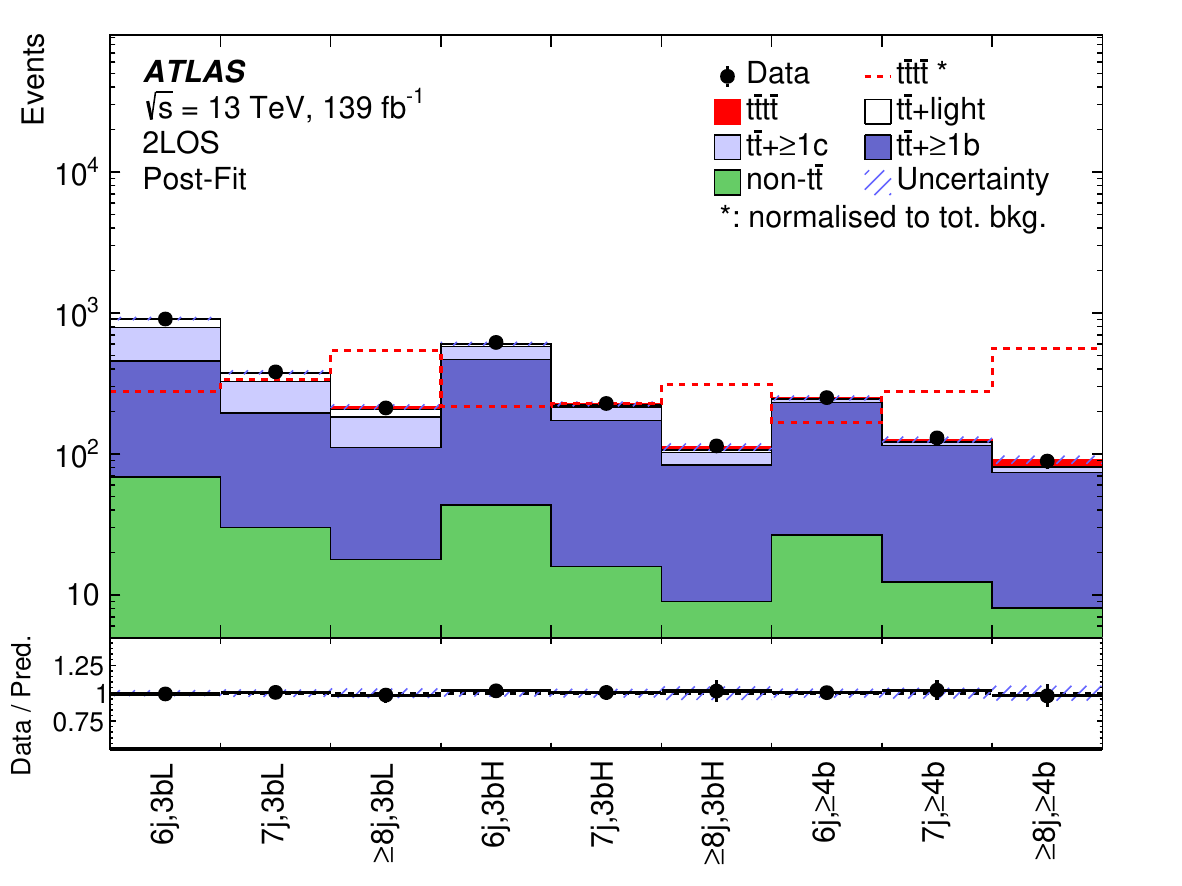}
\caption{
Comparison of predicted and observed event yields in each control and signal region
in the \SL channel (top) and in the \OS channel (bottom) before the fit (left) and after the fit (right).
The \ttjets background is corrected at the pre-fit level using data.
The band includes the total pre- or post-fit uncertainties.
The dashed red line shows the signal distribution normalised to the total background yield.
The ratio of the data to the total pre- or post-fit prediction is shown in the lower panel.
}
\label{fig:summary}
\end{figure}
 
\begin{figure}[!htb]
\centering
\includegraphics[width=0.35\textwidth]{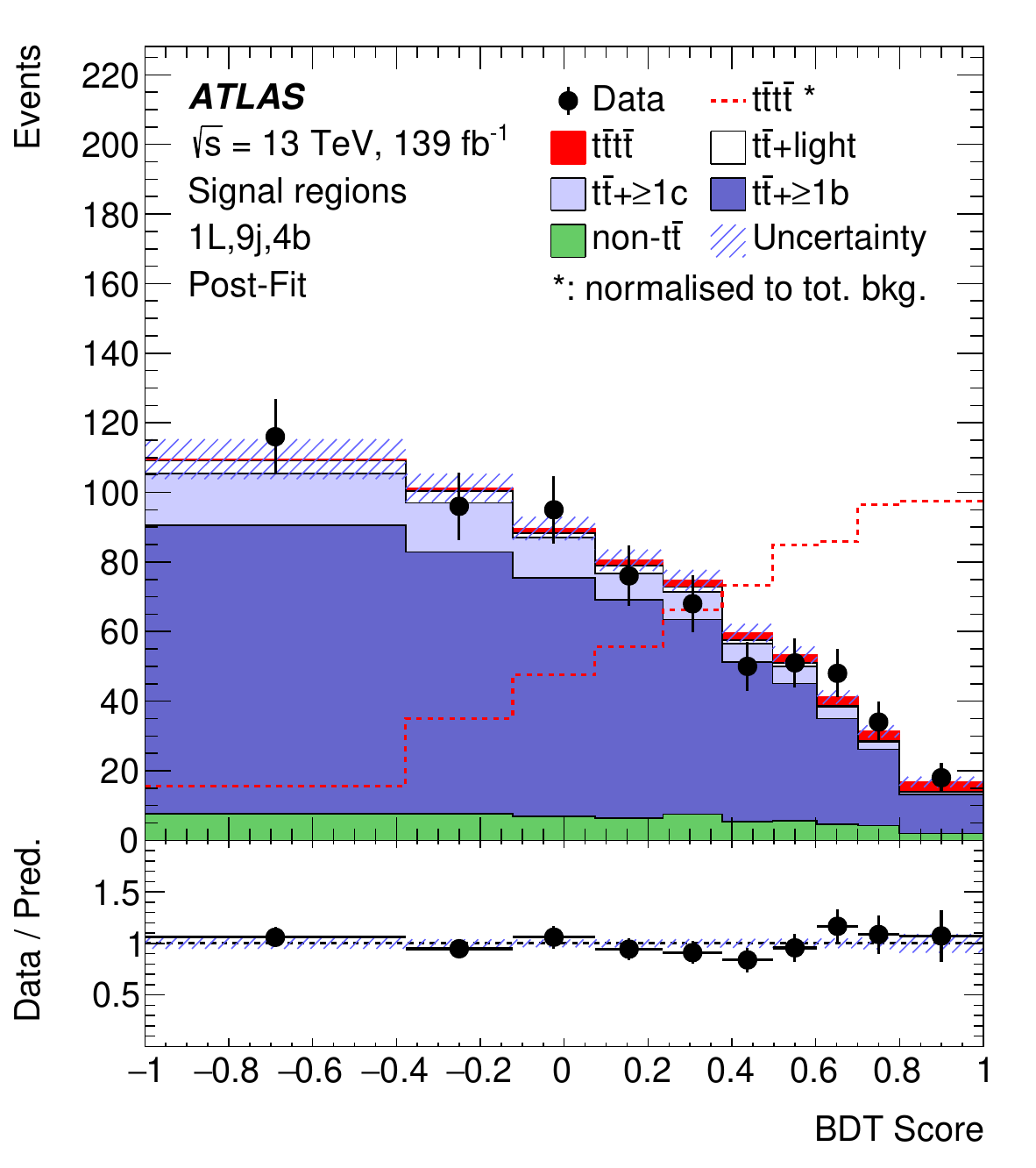}
\includegraphics[width=0.35\textwidth]{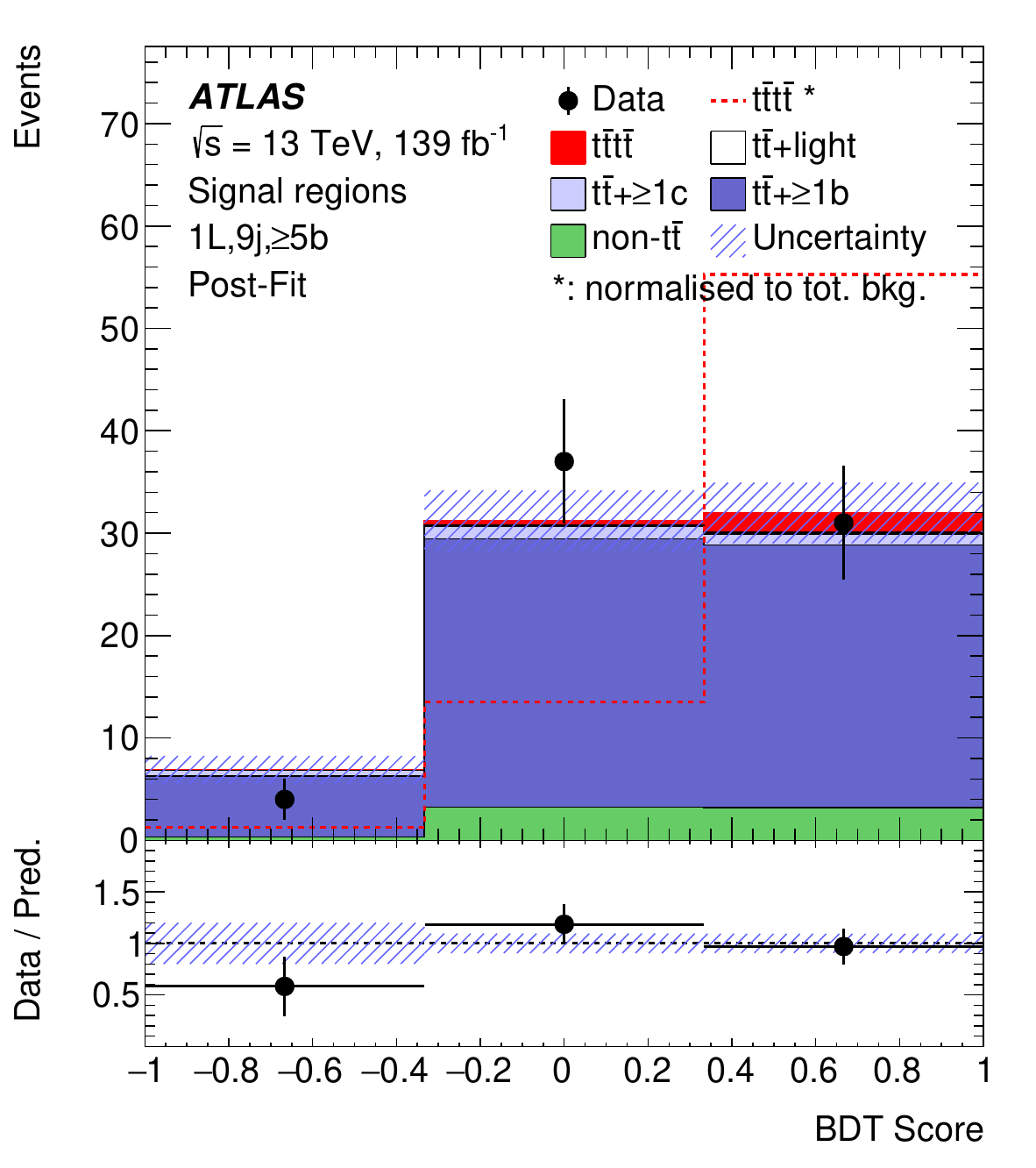}
\includegraphics[width=0.35\textwidth]{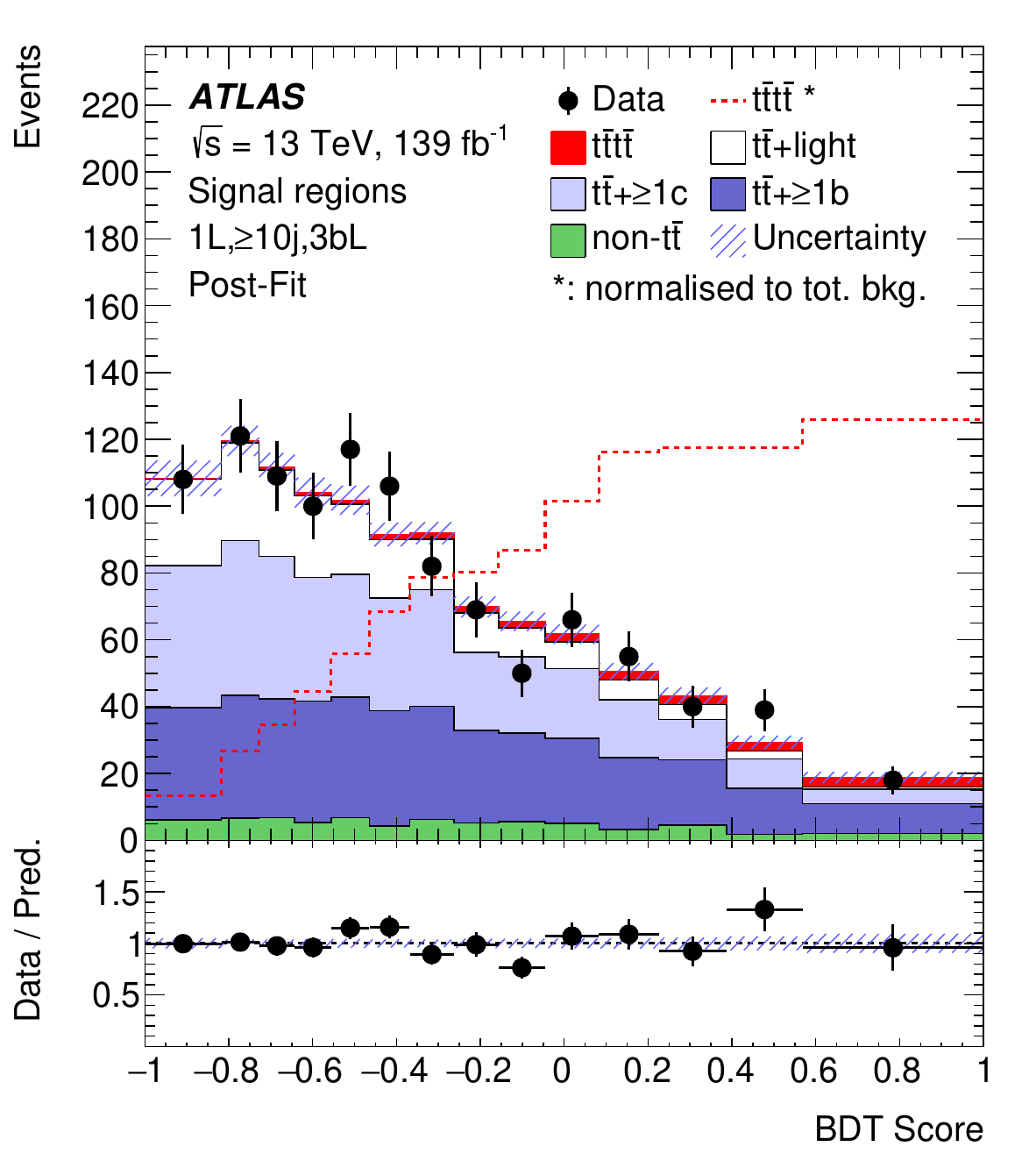}
\includegraphics[width=0.35\textwidth]{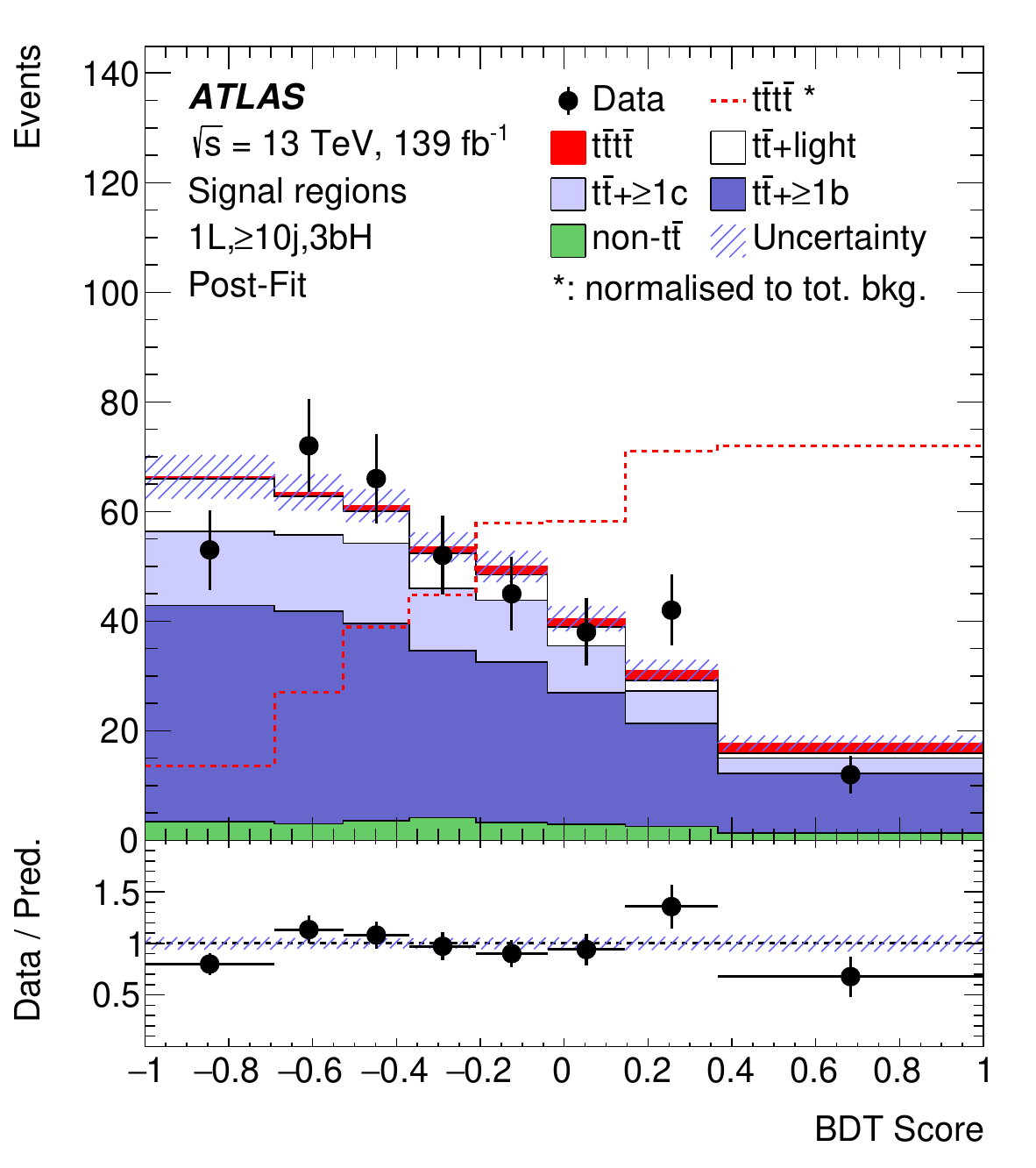}
\includegraphics[width=0.35\textwidth]{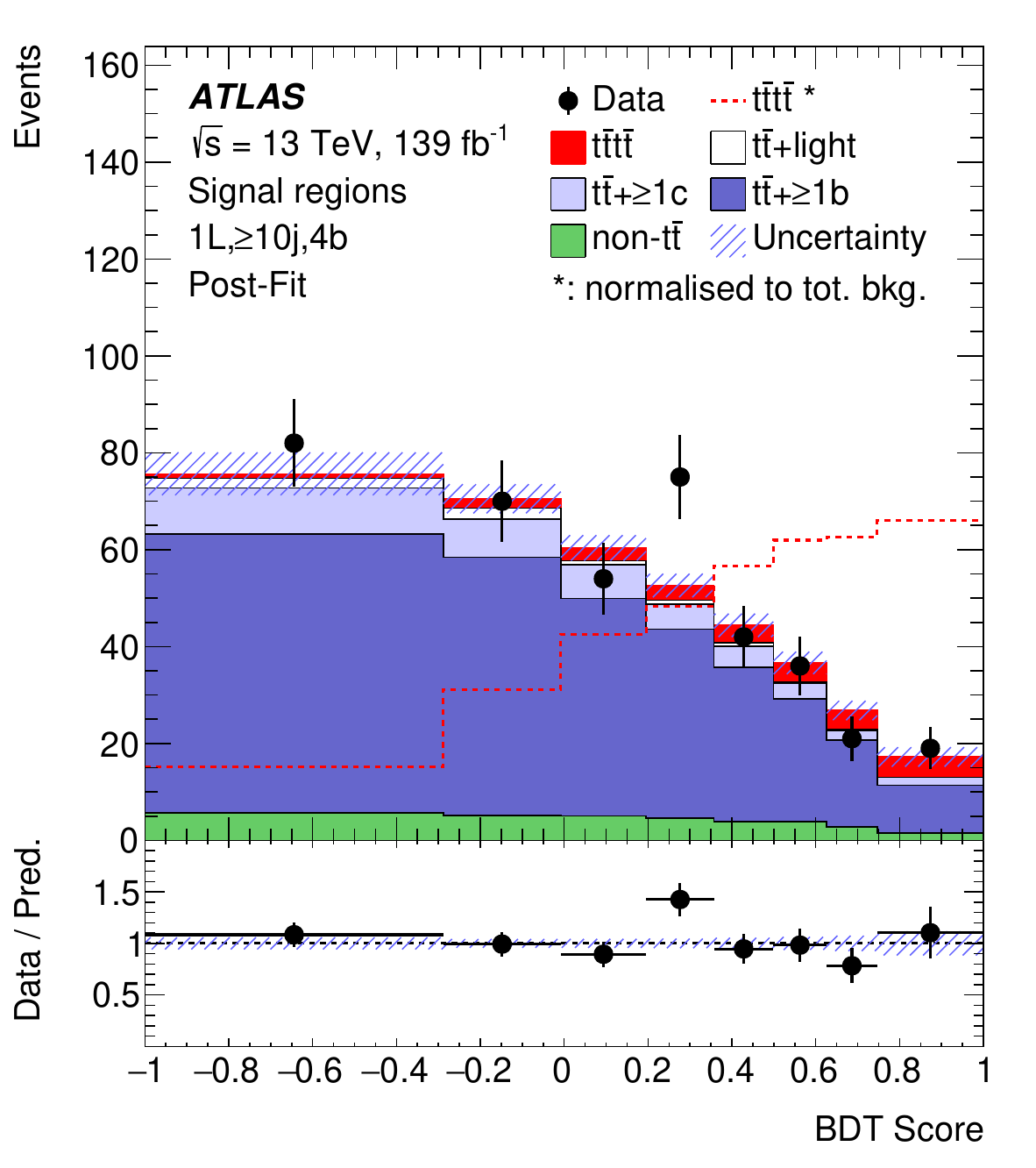}
\includegraphics[width=0.35\textwidth]{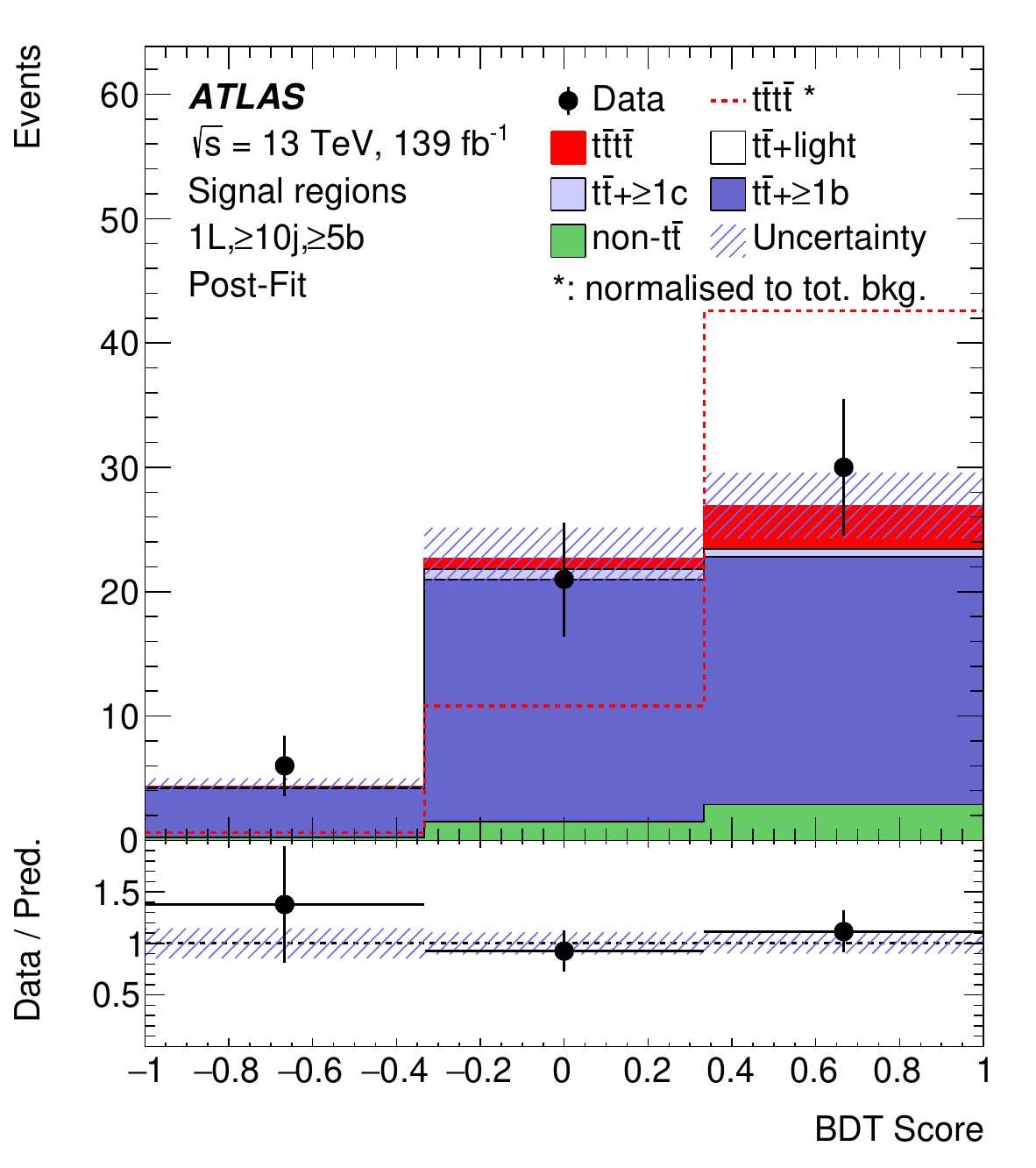}
\caption{
Comparison between data and post-fit prediction for the distributions of the BDT score in each signal region in the \SL channel.
The band includes the total uncertainty of the post-fit computation.
The dashed red line shows the signal distribution normalised to the background yield.
The ratio of the data to the total post-fit computation is shown in the lower panel.
}
\label{fig:sr_postfit_1l}
\end{figure}
 
\begin{figure}[!htb]
\centering
\includegraphics[width=0.45\textwidth]{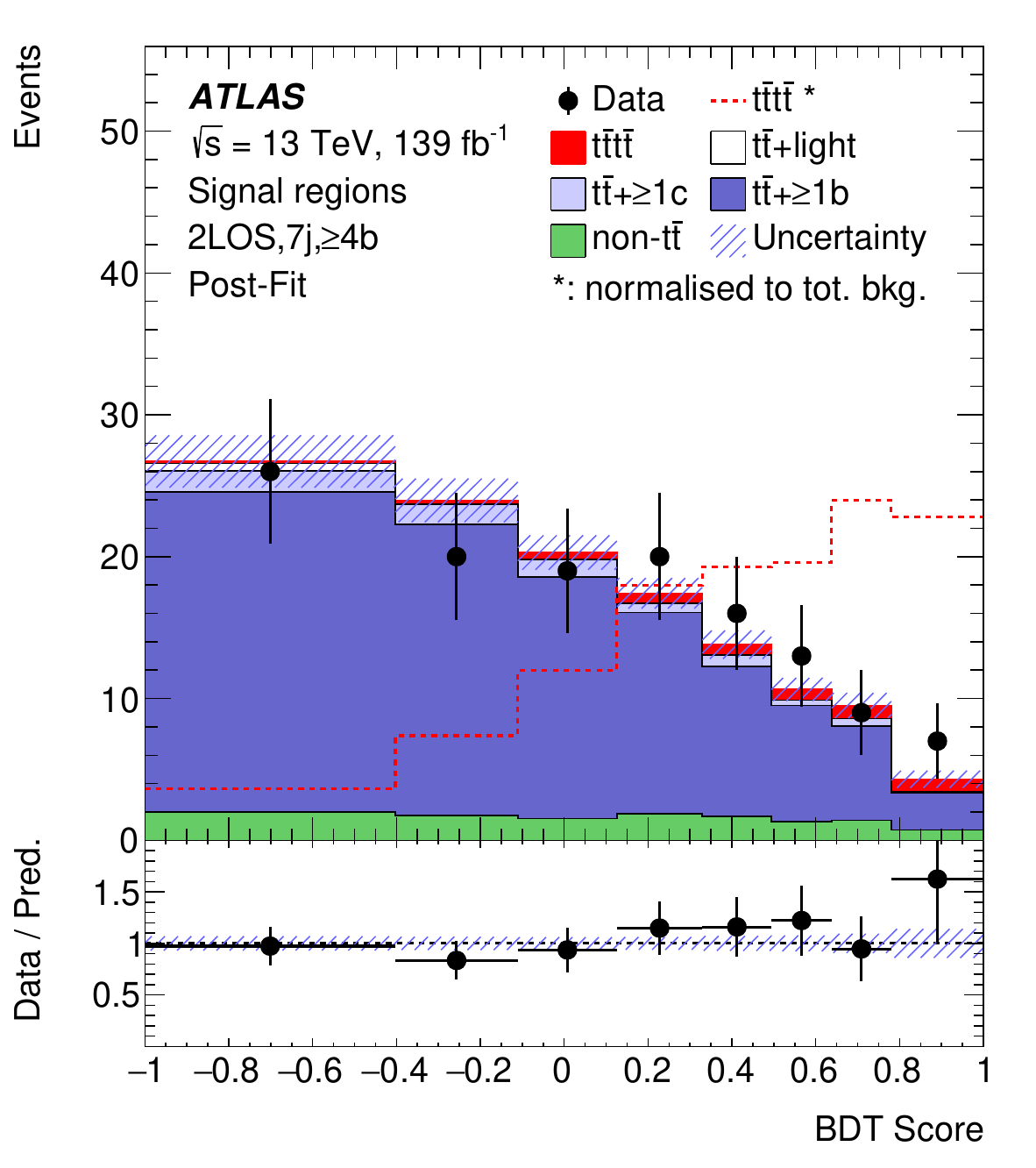}
\includegraphics[width=0.45\textwidth]{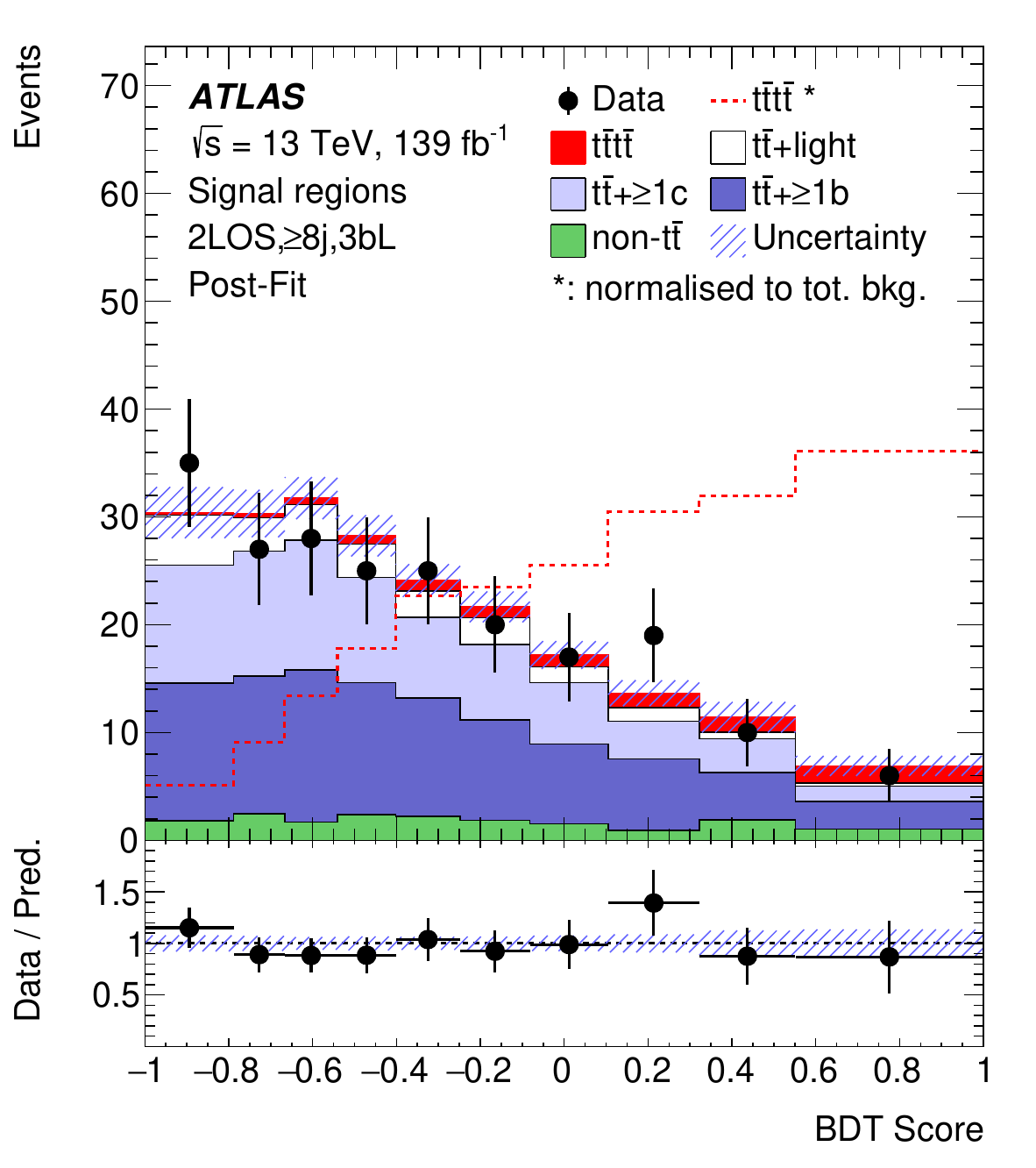}
\includegraphics[width=0.45\textwidth]{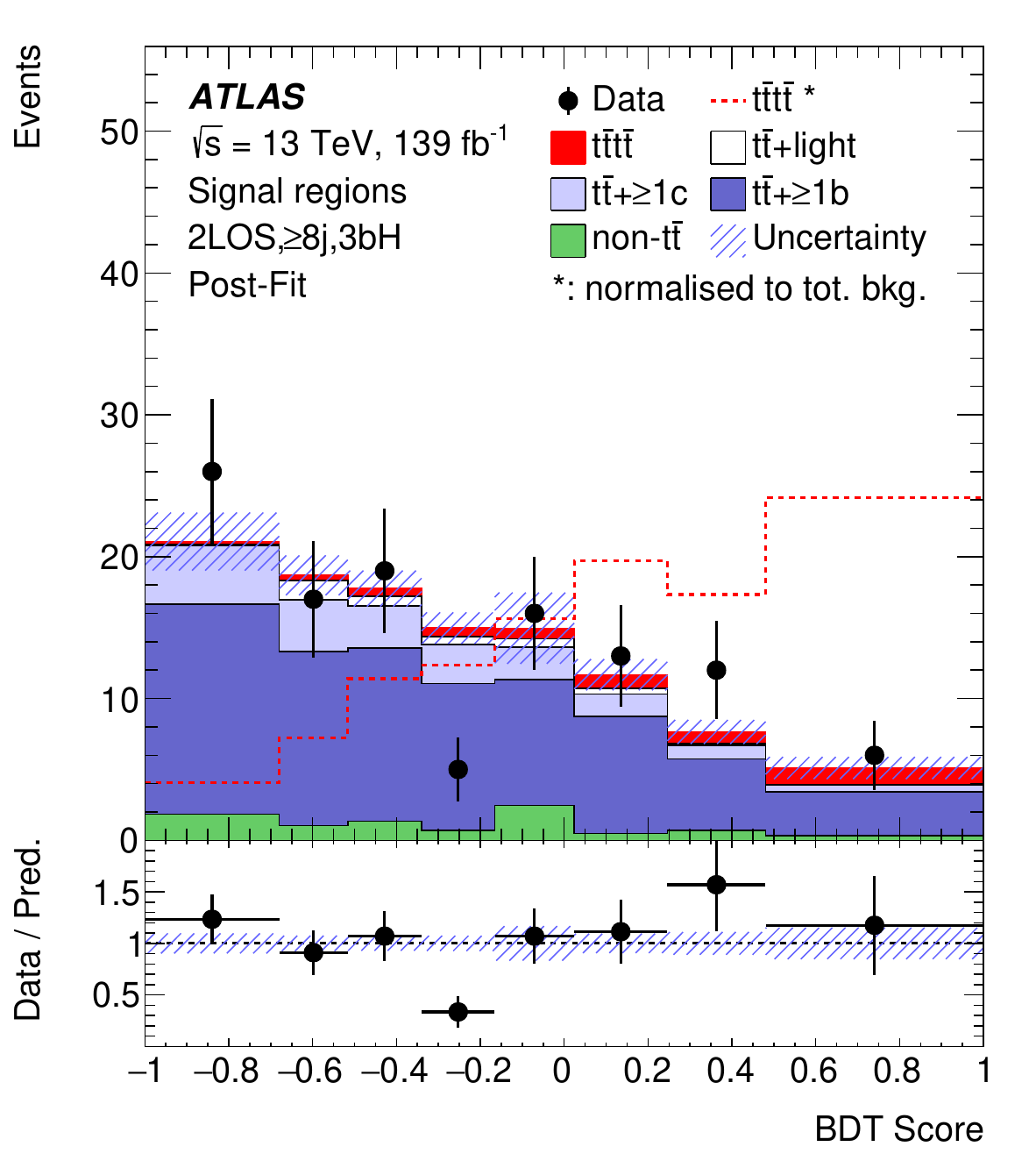}
\includegraphics[width=0.45\textwidth]{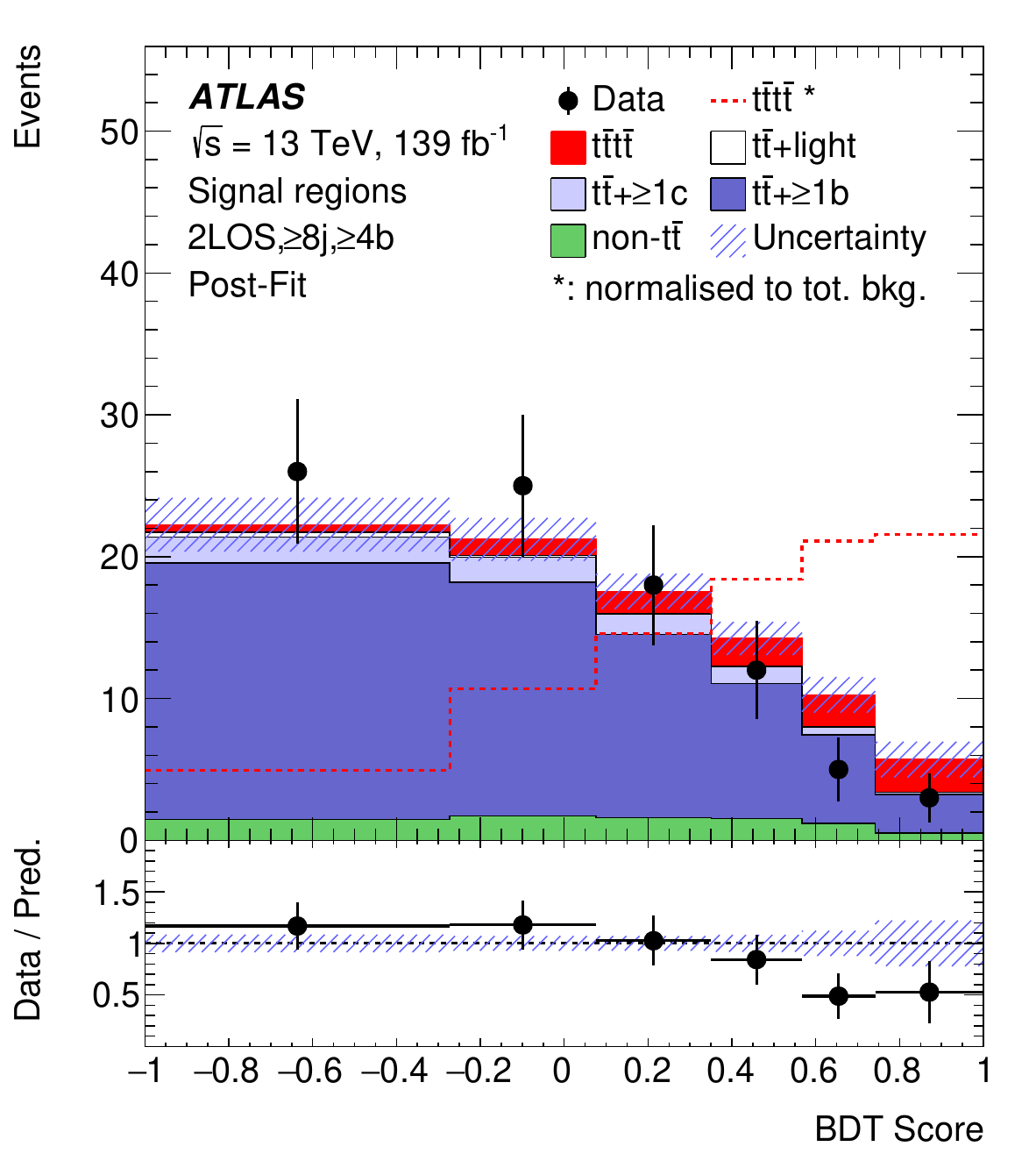}
\caption{
Comparison between data and post-fit prediction for the distributions of the BDT score in each signal region in the \OS channel.
The band includes the total uncertainty of the post-fit computation.
The dashed red line shows the signal distribution normalised to the background yield.
The ratio of the data to the total post-fit computation is shown in the lower panel.
}
\label{fig:sr_postfit_2l}
\end{figure}
 
\begin{figure}[!htb]
\centering
\includegraphics[width=0.45\textwidth]{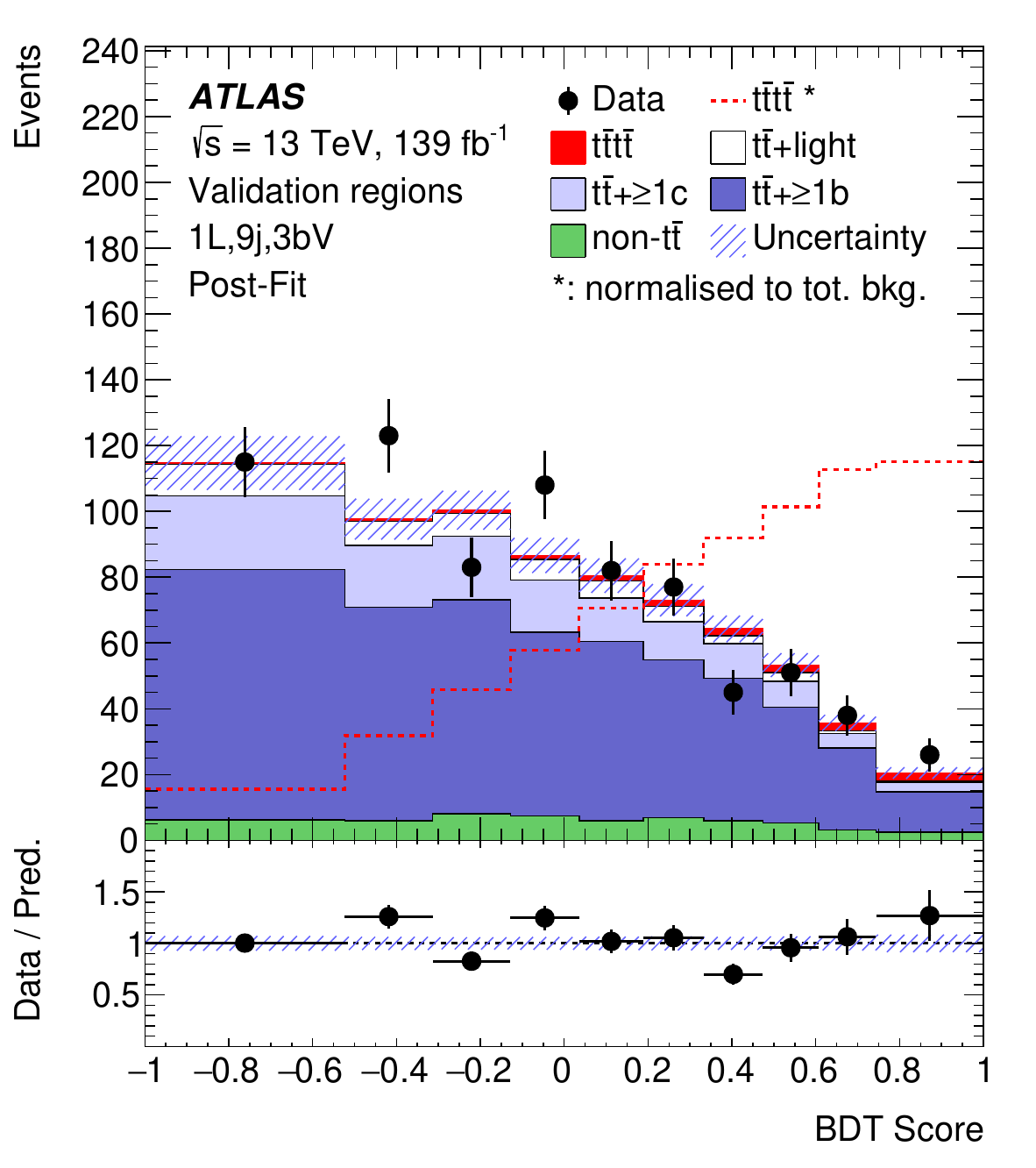}
\includegraphics[width=0.45\textwidth]{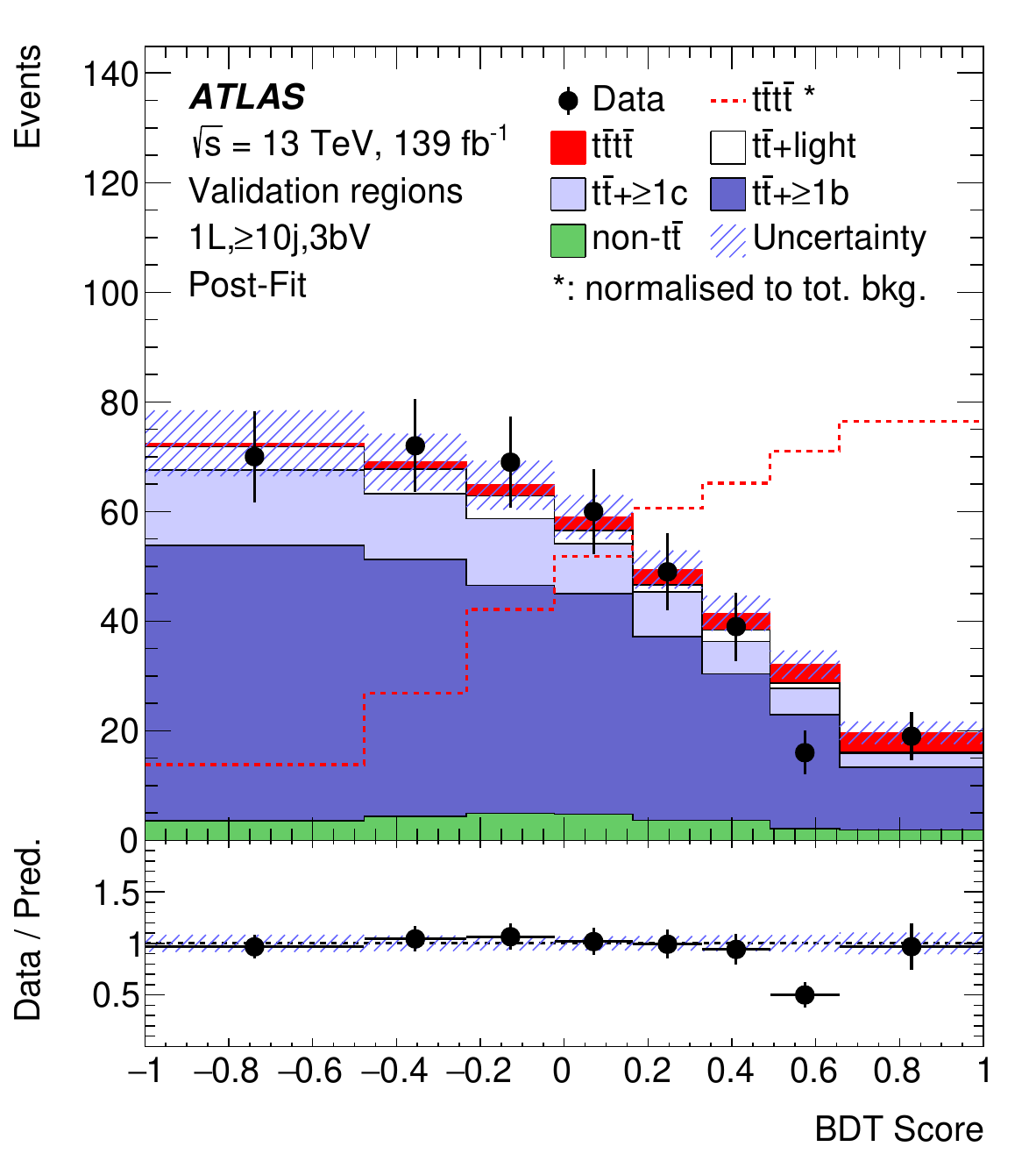}
\includegraphics[width=0.45\textwidth]{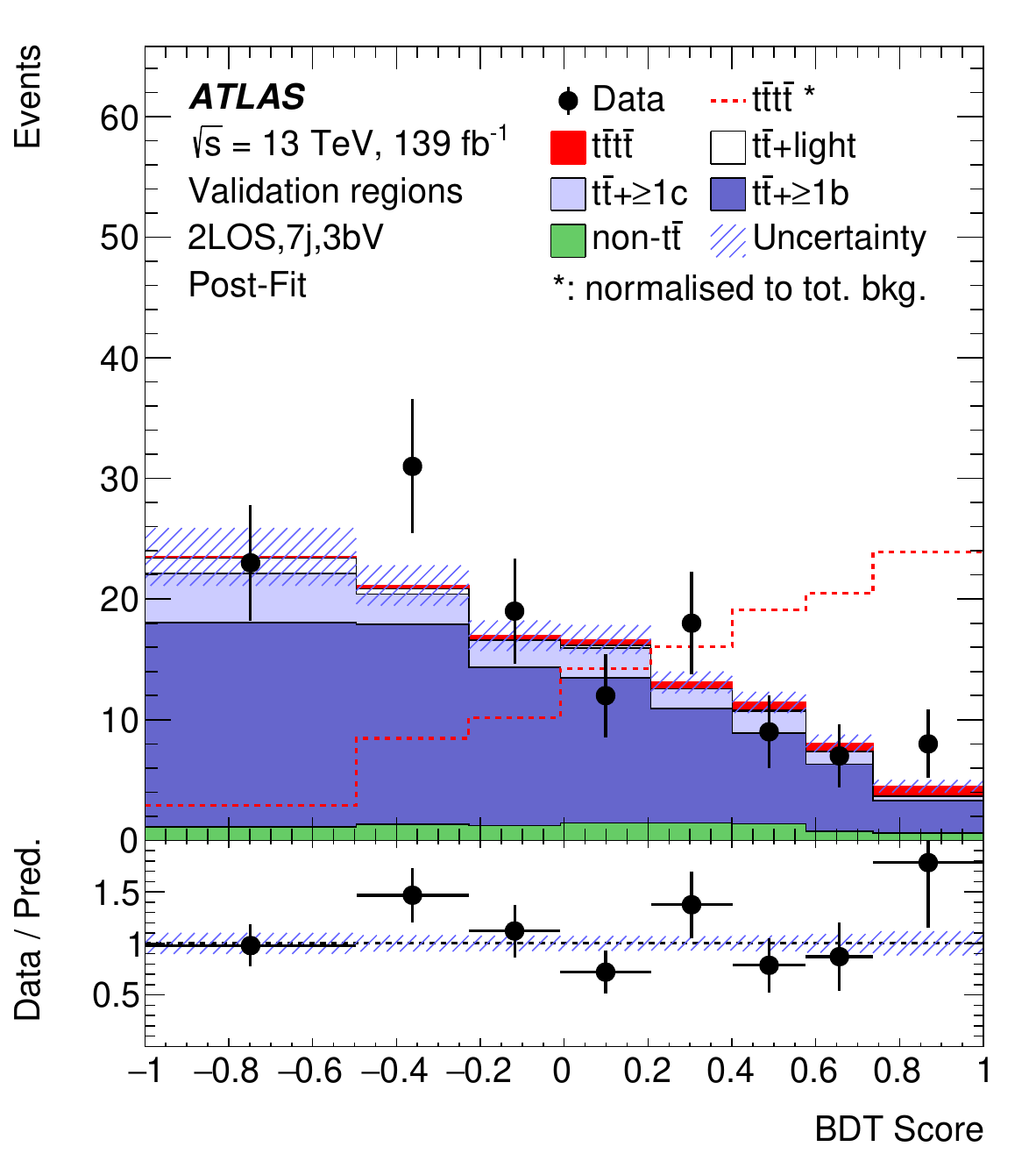}
\includegraphics[width=0.45\textwidth]{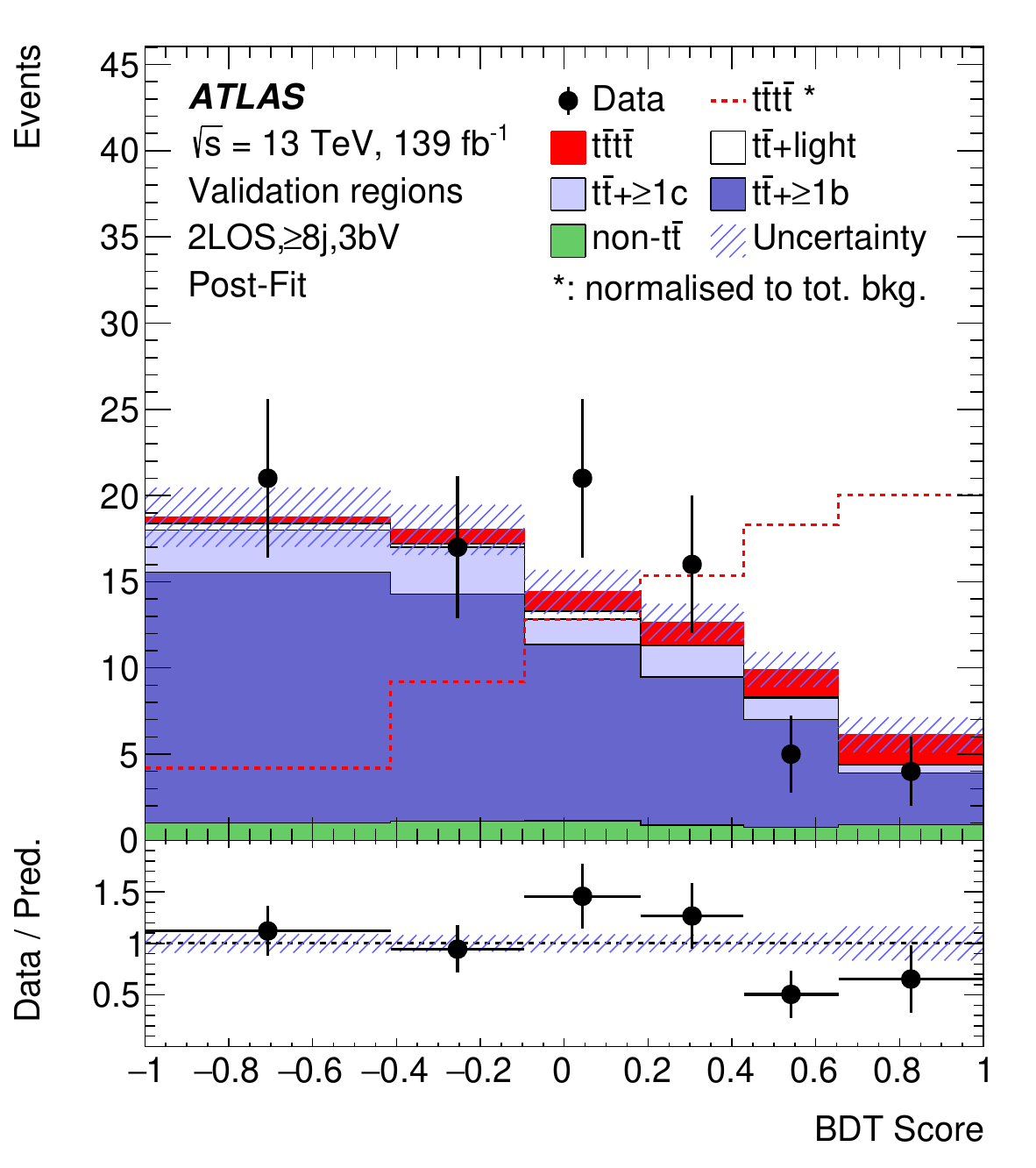}
\caption{
Comparison between data and post-fit prediction for the
distributions of the BDT score in each validation region.
The band includes the total uncertainty of the post-fit computation.
The dashed red line shows the signal distribution normalised to the background yield.
The ratio of the data to the total post-fit computation is shown in the lower panel.
}
\label{fig:vr_postfit}
\end{figure}
 
The best-fit value for the signal strength is
\begin{equation}\nonumber
\mu=2.2 \pm 0.7\,\text{(stat.)}\,^{+1.5}_{-1.0}\,\text{(syst.)}=2.2\,^{+1.6}_{-1.2},
\end{equation}
which takes into account the additional 20\% uncertainty in $\sigma^{\text{SM}}_{\tttt}$ in the
denominator of the $\mu$ parameter.
In order to check the compatibility of the \SL and \OS channels,
an alternative fit is performed with two independent signal-strength parameters,
one for each of the two channels.
The best-fit values of the two signal strengths are $2.9\,^{+1.8}_{-1.5}$ for the \SL channel
and $1.3\,^{+1.7}_{-1.5}$ for the \OS channel.
The probability of obtaining a discrepancy between these two signal-strength parameters equal to or
larger than the one observed is 39\%.
The compatibility between the signal regions was also checked by performing a fit using a different
signal-strength parameter in each region.
The fitted values are found to be consistent within uncertainties.

After excluding the uncertainty in the SM \tttt cross section,
the \tttt\ production cross section is measured to be
\begin{equation}\nonumber
\sigma_{\tttt}= 26 \pm 8\,\text{(stat.)} \,^{+15}_{-13}\,\text{(syst.)}\,\text{fb} = 26\,^{+17}_{-15}\,\text{fb}.
\end{equation}
The statistical uncertainty is obtained from a fit where all nuisance parameters are fixed to their
post-fit values.
The systematic uncertainty is determined by subtracting in quadrature the statistical uncertainty
from the total uncertainty.
The measured cross section is consistent with
the SM predicted cross section of $\sigma^{\text{SM}}_{\tttt}=12.0\pm2.4~\text{fb}$
computed at NLO in QCD including NLO electroweak corrections~\cite{Frederix:2017wme}.
The observed significance of \tttt production relative
to the background-only prediction is 1.9 standard deviations,
while 1.0 standard deviations is expected.
The observed (expected) $p$-value is 3\% (17\%),
corresponding to the probability of obtaining a result
at least as signal-like as observed if no signal is present.

Figure~\ref{fig:soverb_1los} shows the event yield in data compared with
the post-fit signal ($S$) and total background ($B$) prediction.
It is ordered by the signal-to-background ratio of the discriminant bins
from all fitted regions.
The predictions are shown for the best-fit signal strength and for the SM prediction.
The observed data shows a signal-like excess over the background,
with a higher probability of compatibility with the best-fit signal strength of $\mu = 2.2$
than with the SM prediction of $\mu=1.0$ in the high $\log_{10}(S/B)$ region.

\begin{figure}[!p]
\centering
\includegraphics[width=0.65\textwidth]{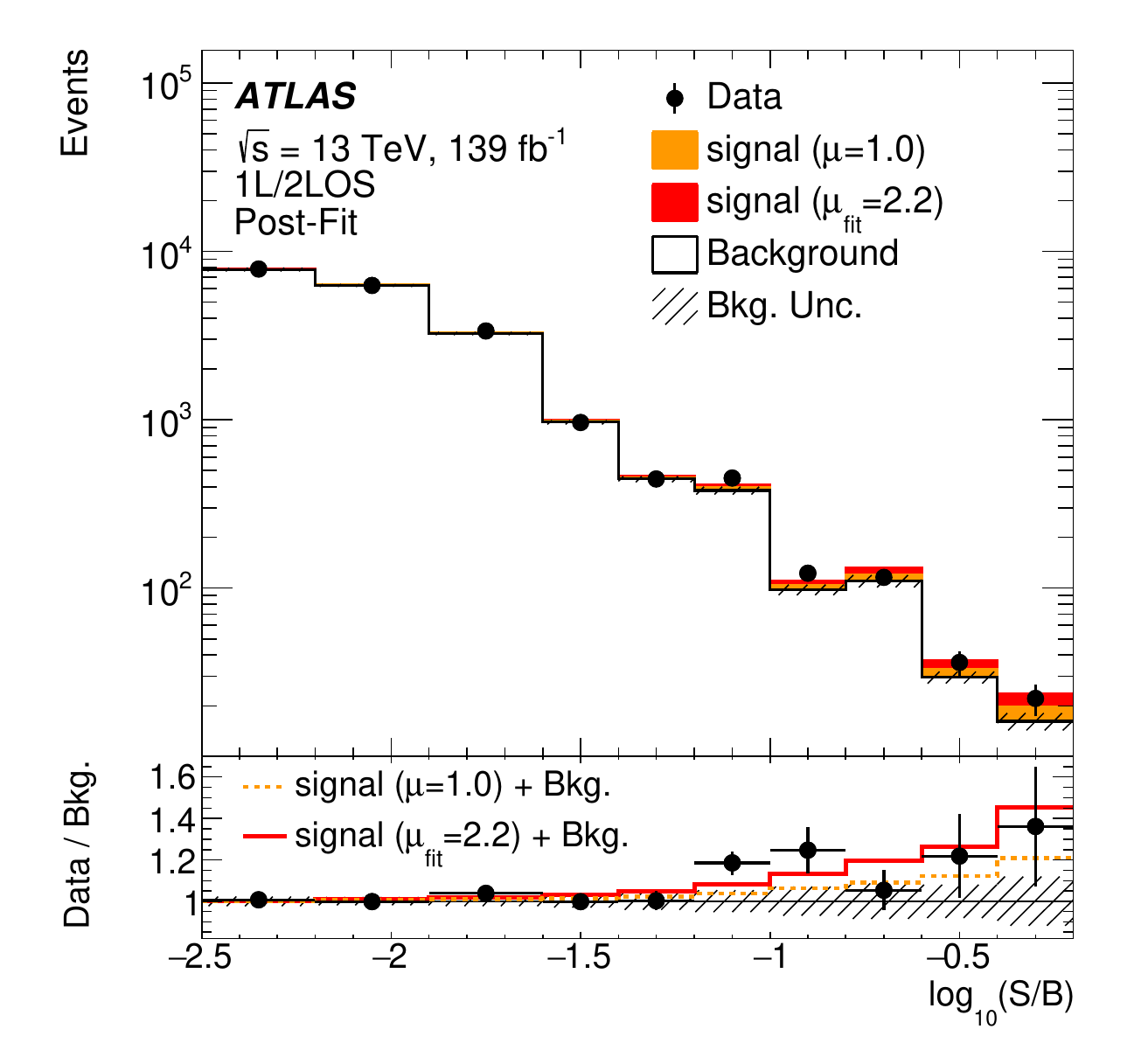}
\caption{
Observed and expected event yields as a function of $\log_{10}(S/B)$,
where $S$ and $B$ are the post-fit signal and total background yields, respectively.
The bins in all fitted regions are ordered and grouped in bins of $\log_{10}(S/B)$.
The signal is shown for both the best-fit signal strength, $\mu=2.2$, and the SM prediction, $\mu=1.0$.
The lower panel shows the ratio of the data to the post-fit background prediction,
compared with the signal-plus-background prediction with the best-fit signal strength
and the SM prediction.
The shaded band represents the total uncertainty in the background prediction.
}
\label{fig:soverb_1los}
\end{figure}

Table~\ref{tab:grouped_syst} lists the contributions from the various groups of systematic uncertainties
to the fitted uncertainty in the measured \tttt production cross section, together with the total systematic uncertainty,
the statistical uncertainty and the total uncertainty.
The impacts of the individual uncertainties on the best-fit value of $\mu$ are presented in Figure~\ref{fig:ranking},
ranked according to their post-fit impacts.
The 20 systematic uncertainties with the largest impact are shown.
The impacts of the individual and grouped systematic uncertainties are evaluated using different methods,
as described in the corresponding captions.

Apart from the uncertainties in the \tttt\ signal,
the largest contribution is from \ttbjets\ modelling uncertainties.
The major uncertainties in this category are due to the choice of flavour scheme and MC generator
used to model the \ttbb background, which is dominant in all signal regions.
This reflects the large difference between the different predictions,
which is expected given our limited knowledge of the phase space being probed.
The contribution from \ttcjets\ modelling uncertainties is dominated by the \ttcjets\ cross-section uncertainty.
Although the fraction of \ttcjets\ events is less than 10\% in the signal regions,
this uncertainty acquires a large anti-correlation with the \ttbb\ cross-section uncertainty
because of the interplay between \ttcjets\ and \ttbb\ in the 3b regions.
The \ttcjets\ cross-section uncertainty is constrained by 50\% due to the substantial impact
of this uncertainty on the regions with loose $b$-tagging requirements,
where abundant data is available to provide the constraint.
The \ttbb\ cross-section uncertainty does not have a large post-fit impact on $\mu$ by itself.
This is because it is constrained by 70\% relative to its pre-fit uncertainty
due to the high purity of \ttbb\ and the sufficient amount of data in the $\geq4$b regions.
The uncertainties in the jet energy scale and resolution also contribute substantially to the overall uncertainty.
This is due to the high \Nj\ multiplicity feature of the \tttt\ signal.
The uncertainties in the $b$-tagging efficiencies are also an important source.
The most important contribution is from `light jets mis-tag rates EV0',
which is the leading component in the 20 uncertainties related to the mis-tag rates of light-flavour jets.
This component mainly affects the migration of the different \ttjets\ components
across the regions defined using different $b$-tagging requirements,
subsequently affecting the \ttbb\ yields in the signal regions,
which are highly correlated with the fitted signal strength.

Figure~\ref{fig:ranking} also illustrates the best-fit values of the nuisance parameters
corresponding to the 20 most important systematic uncertainties.
A number of nuisance parameters are pulled from their nominal value in the likelihood fit.
However, all observed pulls are smaller than 50\% of the corresponding prior uncertainty.
The pull on the `\tttt\ PS choice' is driven by
the discrepancy between data and MC simulation in the \OS\ channel, ($\geq$8j, $\geq$4b) region.
The slight pull on this nuisance parameter reduces the \tttt\ yield in this region by 3\%.
The effects in other fitted regions are either comparable or smaller.
The `light jets mis-tag rates EV0' nuisance parameter is pulled to mitigate the discrepancy between data and prediction
by adjusting the relative normalisation of the different \ttjets\ components across the fitted regions.
All other pulls, including those not shown in Figure~\ref{fig:ranking}, are mainly on \ttjets\ modelling uncertainties.
The combination of these pulls is found to correct for the mismodelling
in the shape and normalisation of the \ttjets\ background prediction.
Similar pulls are observed in a background-only fit when removing the bins with a signal-to-background ratio larger than 5\%.
Additional tests were performed with pseudo-data sets constructed using various alternative \ttjets\ predictions and injected signal.
No bias in the extracted $\mu$ was seen in the fits to these pseudo-data sets.

\begin{table}[!hbt]
\begin{center}
\caption{
The contribution from different systematic uncertainties to the measured \tttt\ production cross section, $\sigma_{\tttt}$, grouped into categories.
For each uncertainty source, the fit is repeated with the corresponding group of nuisance parameters
fixed to their best-fit values, $\hat{\pmb{\theta}}$.
The contribution from each source, $\Delta\sigma_{\tttt}$, is then evaluated
by subtracting in quadrature the uncertainty in $\sigma_{\tttt}$ obtained in this fit from that of the full fit.
The contributions from individual groups are compared with the total systematic uncertainty and the statistical uncertainty.
The total systematic uncertainty is different from the sum in quadrature of the different groups due to correlations
among nuisance parameters in the fit.
\newline
}
\begin{tabular}{lrr}
\hline
\hline
\\[-1em]
Uncertainty source & \multicolumn{2}{c}{$\Delta\sigma_{\tttt}$~[fb]} \\
\\[-1em]
\hline
\\[-1em]
\textbf{Signal Modelling} && \\
\tttt\ modelling                         & $+8$ & $-3$  \\
\\[-1em]
\hline
\\[-1em]
\textbf{Background Modelling} && \\
\ttbjets\ modelling                      & $+8$ & $-7$  \\
\ttcjets\ modelling                      & $+5$ & $-4$  \\
\ttjets\ reweighting                     & $+4$ & $-3$  \\
Other background modelling               & $+4$ & $-3$  \\
\ttljets\ modelling                      & $+2$ & $-2$  \\
\\[-1em]
\hline
\\[-1em]
\textbf{Experimental} && \\
Jet energy scale and resolution          & $+6$ & $-4$  \\
$b$-tagging efficiency and mis-tag rates & $+4$ & $-3$  \\
MC statistical uncertainties             & $+2$ & $-2$  \\
Luminosity                               & \multicolumn{2}{c}{$<1$}  \\
Other uncertainties                      & \multicolumn{2}{c}{$<1$}  \\
\\[-1em]
\hline
\hline
\\[-1em]
\textbf{Total systematic uncertainty}    & $+15$  & $-12$  \\
\\[-1em]
\hline
\hline
\\[-1em]
\textbf{Statistical uncertainty}         & $+8$  & $-8$  \\
\\[-1em]
\hline
\hline
\\[-1em]
\textbf{Total uncertainty}               & $+17$  & $-15$  \\
\\[-1em]
\hline
\hline
\end{tabular}
\label{tab:grouped_syst}
\end{center}
\end{table}

\begin{figure}[!htb]
\centering
\includegraphics[width=0.8\textwidth]{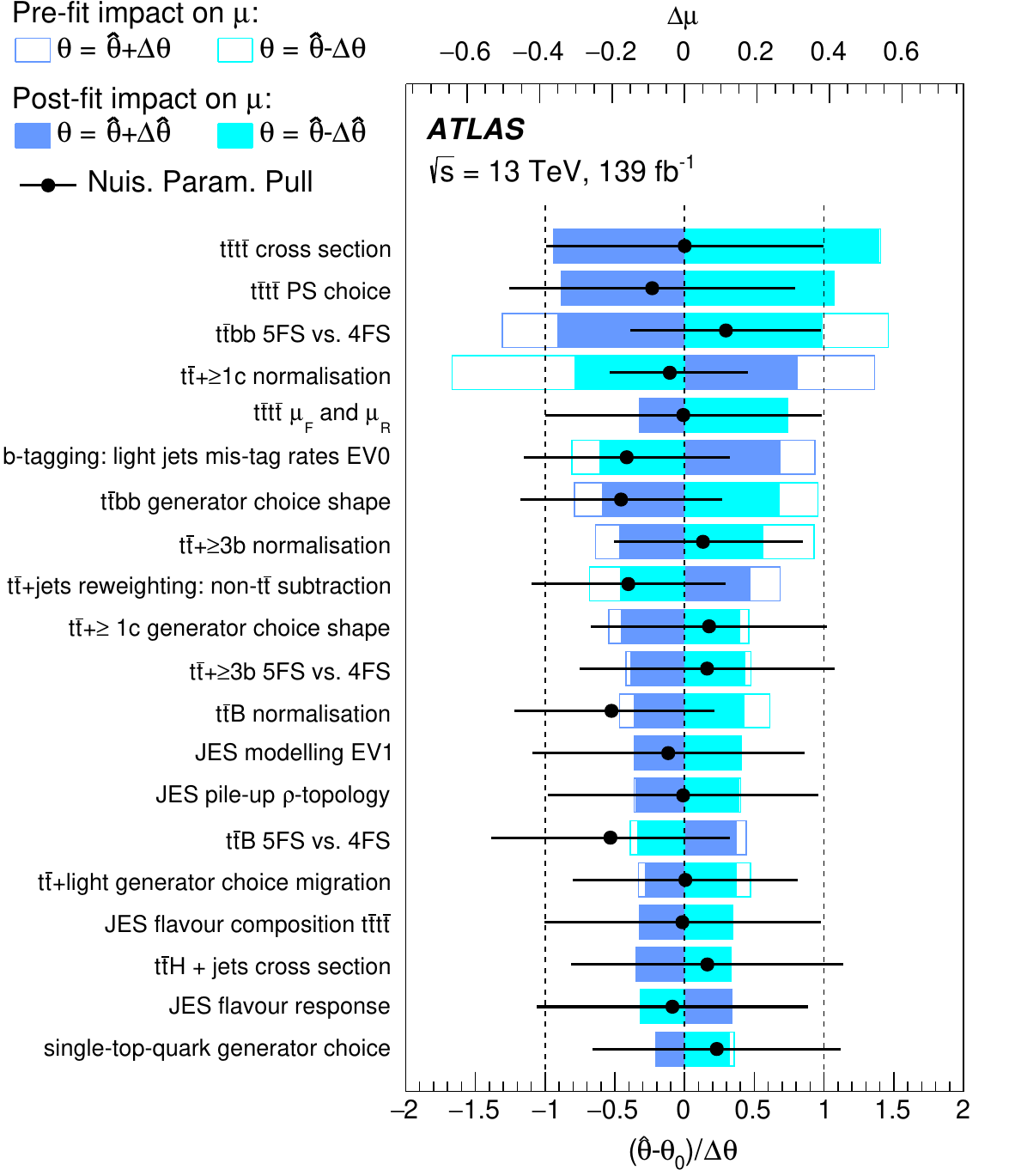}
\caption{
The nuisance parameters ranked according to their post-fit impacts on the best-fit value of $\mu$.
Only the 20 nuisance parameters with the largest impacts are shown.
The empty (solid) blue rectangles illustrate the pre-fit (post-fit) impacts
on the parameter of interest $\mu$, corresponding to the top axis.
The pre-fit (post-fit) impact of each nuisance parameter, $\Delta\mu$,
is calculated as the difference in the fitted value of $\mu$
between the nominal fit and the fit when fixing the corresponding nuisance parameter
to $\hat\theta \pm \Delta\theta$ ($\hat\theta \pm \Delta\hat\theta$),
where $\hat\theta$ is the best-fit value of the nuisance parameter
and $\Delta\theta$ ($\Delta\hat\theta$) is its pre-fit (post-fit) uncertainty.
The black points show the best-fit values of the nuisance parameters,
with the error bars representing the post-fit uncertainties.
Each nuisance parameter is shown relative to its nominal value, $\theta_{0}$,
and in units of its pre-fit uncertainty.
}
\label{fig:ranking}
\end{figure}

% End of text imported from the .//results/results.tex input file
 
% The next lines are included from the .//combination/combination.tex input file
 
\section{Combination with the same-sign dilepton and multilepton final state}
\label{sec:combination}
 
The measurement of \tttt production in the \SLOS\ final state presented in this article
is combined with the ATLAS measurement in the \SSML\ final state~\cite{Aad:2020klt}
using the same data set. The combination is performed via
a simultaneous profile likelihood fit in all regions of both
analyses, with all systematic uncertainties included,
and the combined \tttt\ production cross section is extracted.
The events in the two analyses are statistically independent by construction
due to the different lepton selection criteria.
 
The systematic uncertainties have a significant impact on the sensitivity
of both the \SLOS\ and \SSML\ analyses. However, because the most relevant systematic
uncertainties in these final states are different, the assumptions about the
correlations of the uncertainties have a negligible effect on the
combined result.
 
The uncertainties related to the modelling of the \ttjets\ background
are treated as uncorrelated between the \SLOS\ and \SSML\ final states.
The contribution of \ttjets\ events in the \SSML\ final state is much smaller
than that in the \SLOS final state,
and arises from misreconstructed or non-prompt leptons, while it is
the dominant irreducible background in the \SLOS\ final state.
Its estimation and the treatment of relevant uncertainties
is therefore different in the two final states.
The experimental uncertainties are treated as fully correlated
between the two final states since both analyses use the same
reconstructed objects and data set.
Theoretical modelling uncertainties in the non-\ttbar backgrounds
and the \tttt\ signal are also fully correlated except for the
normalisation of the \ttW background, which is treated differently in the
\SLOS\ and \SSML\ final states. In the latter, \ttW is the most important
background and its normalisation is a free parameter of the
fit, while in the former it is a small background and its normalisation
is taken from MC simulation with a large uncertainty motivated by the normalisation
factor obtained in the \SSML\ analysis. For this reason, the
\ttW normalisation is uncorrelated between the two final states.
The systematic uncertainties associated with the reducible backgrounds in the \SSML\ final state
and uncertainties related to the data-driven corrections to the \ttjets background in the \SLOS\ final state
are uncorrelated.

The \tttt\ signal strength from the combination of the two final states is measured to be
\begin{equation}\nonumber
\mu=2.0 \pm 0.4\,\text{(stat.)}\,^{+0.7}_{-0.5}\,\text{(syst.)}=2.0\,^{+0.8}_{-0.6}.
\end{equation}
The best-fit \tttt\ signal strengths
for the \SLOS\ and \SSML\ final states fitted separately,
and for the combination of the two final states,
are summarised in Figure~\ref{fig:comb:poi}.
The fitted $\mu$ values are consistent between the two final states,
which are dominated by different backgrounds and
different uncertainties.
The combined SM \tttt\ production cross section is measured to be
\begin{equation}\nonumber
\sigma_{\tttt} = 24 \pm 4\,\text{(stat.)}\,^{+5}_{-4}\,\text{(syst.)}\,\text{fb} = 24\,^{+7}_{-6}\,\text{fb}.
\end{equation}
The measured cross section is consistent within 2.0 standard deviations with
the SM prediction of $\sigma^{\text{SM}}_{\tttt}=12.0\pm2.4~\text{fb}$
computed at NLO in QCD including NLO electroweak corrections~\cite{Frederix:2017wme}.
The observed (expected) significance of the combined measured \tttt\ signal
is $4.7\sigma$~($2.6\sigma$) above the background-only hypothesis.
The significance is improved compared to the observed (expected) significance of $4.3\sigma$ ($2.4\sigma$) obtained in the \SSML\ analysis.

\begin{figure}[!p]
\centering
\includegraphics[width=0.6\textwidth]{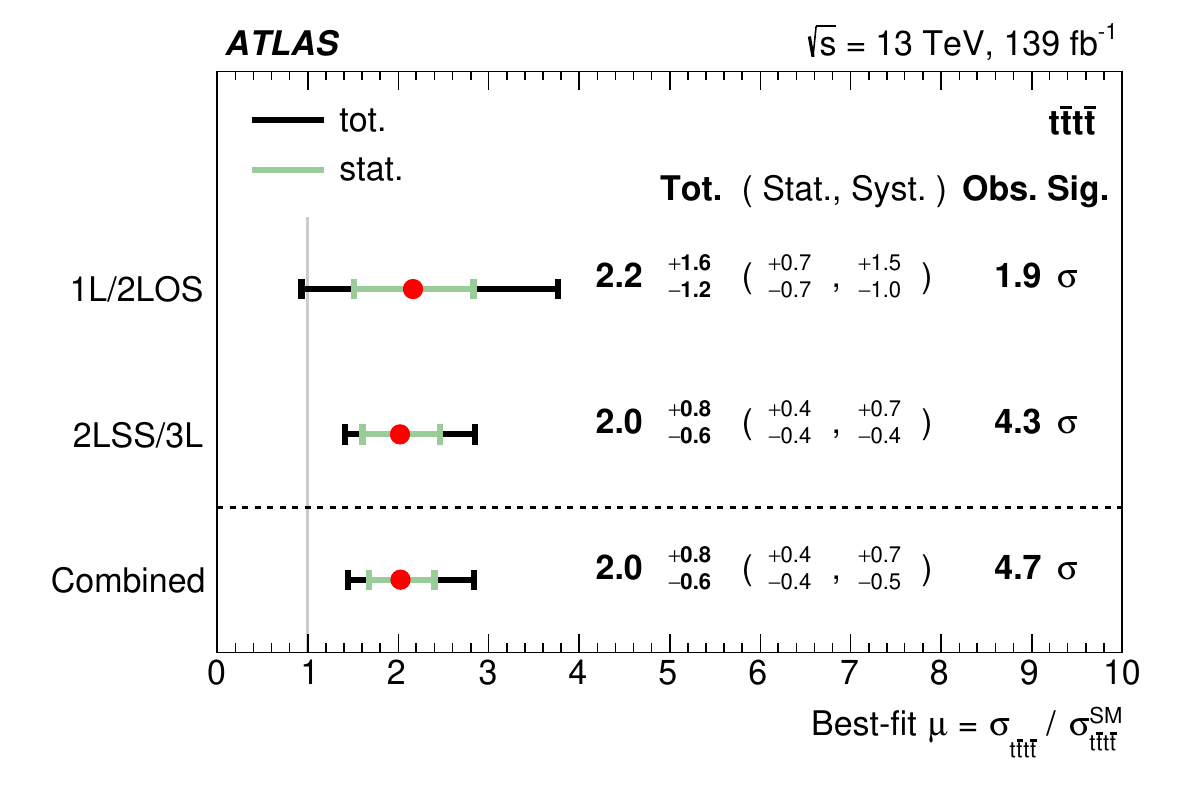}
\caption{Fitted signal strengths in the signal-plus-background fits to the same
data set for the individual final states (\SLOS and \SSML) and the combined signal strength from all \tttt\ analysis regions.
}
\label{fig:comb:poi}
\end{figure}

% End of text imported from the .//combination/combination.tex input file
 
\FloatBarrier
 
% The next lines are included from the .//conclusion/conclusion.tex input file
\section{Conclusion}
\label{sec:conclusion}
A measurement of four-top-quark production using 139~fb\(^{-1}\) of $\sqrt{s}=13~\tev$
proton--proton collision data collected by the ATLAS detector at the Large Hadron Collider
is presented.
 
The selection in the \SLOS\ final state targets events with
a single lepton or an opposite-sign lepton pair accompanied by multiple jets.
The measured four-top-quark production cross section is found to be
$26 \pm 8 \,\text{(stat.)} \, ^{+15}_{-13} \, \text{(syst.)}$~fb, corresponding to an observed (expected) signal
significance of 1.9 (1.0) standard
deviations over the background-only hypothesis. The measurement uncertainty
is dominated by the
systematic uncertainties, particularly the modelling uncertainty
of the \tttt signal and the \ttHF background.
 
The result in the \SLOS\ final state is combined with the previous measurement
performed by the ATLAS Collaboration in the multilepton final state with the
same data set.
The combined four-top-quark production cross section is measured to
be $\sigma_{\tttt}=24 \pm 4 \,\text{(stat.)} \, ^{+5}_{-4} \,\text{(syst.)}$~fb.
It is consistent within 2.0 standard deviations with the Standard Model
expectation of $\sigma^{\text{SM}}_{\tttt}=12.0\pm2.4~\text{fb}$.
The observed (expected) significance of the combined \tttt\ cross section
is 4.7~(2.6) standard deviations above the background-only hypothesis.
It improves upon the observed (expected) significance of 4.3 (2.4) standard deviations
from the analysis in the multilepton final state.

% End of text imported from the .//conclusion/conclusion.tex input file

\section*{Acknowledgements}
 
% The next lines are included from the .//acknowledgements/Acknowledgements.tex input file

We thank CERN for the very successful operation of the LHC, as well as the
support staff from our institutions without whom ATLAS could not be
operated efficiently.
 
We acknowledge the support of
ANPCyT, Argentina;
YerPhI, Armenia;
ARC, Australia;
BMWFW and FWF, Austria;
ANAS, Azerbaijan;
SSTC, Belarus;
CNPq and FAPESP, Brazil;
NSERC, NRC and CFI, Canada;
CERN;
ANID, Chile;
CAS, MOST and NSFC, China;
Minciencias, Colombia;
MSMT CR, MPO CR and VSC CR, Czech Republic;
DNRF and DNSRC, Denmark;
IN2P3-CNRS and CEA-DRF/IRFU, France;
SRNSFG, Georgia;
BMBF, HGF and MPG, Germany;
GSRI, Greece;
RGC and Hong Kong SAR, China;
ISF and Benoziyo Center, Israel;
INFN, Italy;
MEXT and JSPS, Japan;
CNRST, Morocco;
NWO, Netherlands;
RCN, Norway;
MNiSW and NCN, Poland;
FCT, Portugal;
MNE/IFA, Romania;
JINR;
MES of Russia and NRC KI, Russian Federation;
MESTD, Serbia;
MSSR, Slovakia;
ARRS and MIZ\v{S}, Slovenia;
DSI/NRF, South Africa;
MICINN, Spain;
SRC and Wallenberg Foundation, Sweden;
SERI, SNSF and Cantons of Bern and Geneva, Switzerland;
MOST, Taiwan;
TAEK, Turkey;
STFC, United Kingdom;
DOE and NSF, United States of America.
In addition, individual groups and members have received support from
BCKDF, CANARIE, Compute Canada and CRC, Canada;
COST, ERC, ERDF, Horizon 2020 and Marie Sk{\l}odowska-Curie Actions, European Union;
Investissements d'Avenir Labex, Investissements d'Avenir Idex and ANR, France;
DFG and AvH Foundation, Germany;
Herakleitos, Thales and Aristeia programmes co-financed by EU-ESF and the Greek NSRF, Greece;
BSF-NSF and GIF, Israel;
Norwegian Financial Mechanism 2014-2021, Norway;
La Caixa Banking Foundation, CERCA Programme Generalitat de Catalunya and PROMETEO and GenT Programmes Generalitat Valenciana, Spain;
G\"{o}ran Gustafssons Stiftelse, Sweden;
The Royal Society and Leverhulme Trust, United Kingdom.
 
The crucial computing support from all WLCG partners is acknowledged gratefully, in particular from CERN, the ATLAS Tier-1 facilities at TRIUMF (Canada), NDGF (Denmark, Norway, Sweden), CC-IN2P3 (France), KIT/GridKA (Germany), INFN-CNAF (Italy), NL-T1 (Netherlands), PIC (Spain), ASGC (Taiwan), RAL (UK) and BNL (USA), the Tier-2 facilities worldwide and large non-WLCG resource providers. Major contributors of computing resources are listed in Ref.~\cite{ATL-SOFT-PUB-2021-003}.
 
% End of text imported from the .//acknowledgements/Acknowledgements.tex input file

\printbibliography
 
\clearpage
\input{atlas_authlist}

\end{document}

%% file: atlas_authlist.tex
% ATLAS Collaboration author list
% Reference date of TOPQ-2020-10 is 2021-03-04
% Author list last updated on date 24-OCT-22
% Data extracted on 24-Oct-2022 for paper reference TOPQ-2020-10
% at 9:07am
 
\begin{flushleft}
\hypersetup{urlcolor=black}
{\Large The ATLAS Collaboration}

\bigskip

\AtlasOrcid[0000-0002-6665-4934]{G.~Aad}$^\textrm{\scriptsize 99}$,    
\AtlasOrcid[0000-0002-5888-2734]{B.~Abbott}$^\textrm{\scriptsize 124}$,    
\AtlasOrcid[0000-0002-7248-3203]{D.C.~Abbott}$^\textrm{\scriptsize 100}$,    
\AtlasOrcid[0000-0002-2788-3822]{A.~Abed~Abud}$^\textrm{\scriptsize 34}$,    
\AtlasOrcid[0000-0002-1002-1652]{K.~Abeling}$^\textrm{\scriptsize 51}$,    
\AtlasOrcid[0000-0002-2987-4006]{D.K.~Abhayasinghe}$^\textrm{\scriptsize 91}$,    
\AtlasOrcid[0000-0002-8496-9294]{S.H.~Abidi}$^\textrm{\scriptsize 27}$,    
\AtlasOrcid[0000-0001-5329-6640]{H.~Abramowicz}$^\textrm{\scriptsize 157}$,    
\AtlasOrcid[0000-0002-1599-2896]{H.~Abreu}$^\textrm{\scriptsize 156}$,    
\AtlasOrcid[0000-0003-0403-3697]{Y.~Abulaiti}$^\textrm{\scriptsize 5}$
% ...
\bigskip
\\

% ...
$^{5}$High Energy Physics Division, Argonne National Laboratory, Argonne IL; United States of America.\\
% ...
$^{27}$Physics Department, Brookhaven National Laboratory, Upton NY; United States of America.\\
% ...
$^{34}$CERN, Geneva; Switzerland.\\
% ...
$^{51}$II. Physikalisches Institut, Georg-August-Universit\"{a}t G\"ottingen, G\"ottingen; Germany.\\
% ...
$^{91}$Department of Physics, Royal Holloway University of London, Egham; United Kingdom.\\
% ...
$^{99}$CPPM, Aix-Marseille Universit\'e, CNRS/IN2P3, Marseille; France.\\
$^{100}$Department of Physics, University of Massachusetts, Amherst MA; United States of America.\\
% ...
$^{124}$Homer L. Dodge Department of Physics and Astronomy, University of Oklahoma, Norman OK; United States of America.\\
% ...
$^{156}$Department of Physics, Technion, Israel Institute of Technology, Haifa; Israel.\\
$^{157}$Raymond and Beverly Sackler School of Physics and Astronomy, Tel Aviv University, Tel Aviv; Israel.\\
% ...

\end{flushleft}

% Created with Glance <Atlas.Glance@cern.ch>